\documentclass[journal,10pt,final,onecolumn]{IEEEtran}
\usepackage[applemac]{inputenc}
\usepackage{amssymb}
\usepackage{graphicx}
\usepackage{xcolor}
\usepackage{multirow}
\usepackage[tight,raggedright]{subfigure}
\usepackage{url}
\usepackage{verbatim} 
\colorlet{dark}{black!}%{red!100!blue!90!black!}%
\urldef{\mailsa}\path|{mohammad.soleymani,thierry.pun}@unige.ch| 
\urldef{\mailsb}\path|{m.a.larson,a.hanjalic}@tudelft.nl| 
\begin{document}
%\title{Corpus Development for Evaluation of Affective Video Indexing}
\title{Corpus Development for Affective Video Indexing}
%\IEEEcompsocitemizethanks is a very special \thanks that produces the bulleted

\author{Mohammad~Soleymani,
        Martha~Larson,
        Thierry~Pun,
        Alan~Hanjalic
\IEEEcompsocitemizethanks{
\IEEEcompsocthanksitem Mohammad Soleymani is with the Intelligent Behaviour Understanding Group (iBUG), Imperial College London, 180 Queen's Gate, London SW7 2AZ, United Kingdom. (Email: m.soleymani@imperial.ac.uk).\protect
\IEEEcompsocthanksitem Martha Larson and Alan Hanjalic are with the Multimedia Information Retrieval Lab, Delft University of Technology, Mekelweg 4, 2628 CD Delft, The Netherlands. (Email: \{m.a.larson,a.hanjalic\}@tudelft.nl).\protect
\IEEEcompsocthanksitem Thierry Pun is with the Computer Vision and Multimedia Laboratory, University of Geneva, Battelle Campus, Building A, Rte. de Drize 7, Carouge(GE) CH - 1227, Switzerland. (Email: thierry.pun@unige.ch).\protect
\IEEEcompsocthanksitem This is a pre-print and longer version of the article which appeared in IEEE Transactions on Multimedia. Please take note, that this version does not completely match the published version.}
\thanks{}}


\markboth{Pre-print version}%
{Shell \MakeLowercase{\textit{et al.}}: Bare Demo of IEEEtran.cls for Signal Processing Society}

%\IEEEpubid{\makebox[\columnwidth]{\hfill 0000--0000/00/\$00.00~\copyright~2007 IEEE}%

% PRIOR to the title within the \IEEEcompsoctitleabstractindextext IEEEtran
\IEEEcompsoctitleabstractindextext{%
\begin{abstract}

Affective video indexing is the area of research that develops techniques to automatically generate descriptions of video content that encode the emotional reactions which the video content evokes in viewers. This paper provides a set of corpus development guidelines based on state-of-the-art practice intended to support researchers in this field. Affective descriptions can be used for video search and browsing systems offering users affective perspectives. The paper is motivated by the observation that affective video indexing has yet to fully profit from the standard corpora (data sets) that have benefited conventional forms of video indexing. Affective video indexing faces unique challenges, since viewer-reported affective reactions are difficult to assess. Moreover affect assessment efforts must be carefully designed in order to both cover the types of affective responses that video content evokes in viewers and also capture the stable and consistent aspects of these responses. We first present background information on affect and multimedia and related work on affective multimedia indexing, including existing corpora. Three dimensions emerge as critical for affective video corpora, and form the basis for our proposed guidelines: the context of viewer response, personal variation among viewers, and the effectiveness and efficiency of corpus creation. Finally, we present examples of three recent corpora and discuss how these corpora make progressive steps towards fulfilling the guidelines.

\end{abstract}
% not want either math or citations to appear in the abstract.

% Note that keywords are not normally used for peer review papers.
\begin{keywords}
Emotional characterization, benchmarks, multimedia, content analysis, videos
\end{keywords}}

\maketitle

% is not selected <OR> if conference mode is selected - because compsoc
% conference papers position the abstract like regular (non-compsoc)
\IEEEdisplaynotcompsoctitleabstractindextext

% \fi
\IEEEpeerreviewmaketitle

\section{Introduction}
\label{sec:intro}
%\IEEEPARstart{T}{his} is the opening sentence of the paper\cite{citeulike:8222108}
%Motivation: Importance of automatic identification/detection of non-topical characteristics of video for recommendation and retrieval
%AH: Fragmented research addressing a wide scope of different tasks -- results are fragmented and not reproducible. 
%What is the relationship between affective annotation and MIR?
%Why are good tasks/corpora necessary?
%Relationship to other types of annotation (TRECVid)

%%%What is affective video indexing? What is it used for?%%%
\IEEEPARstart{V}{ideo} indexing is the process of analyzing video content in order to extract a representation that is specific enough to characterize the uniqueness of the content and, at the same time, is abstract enough to capture useful similarities with other video content. Research and development in the area of video indexing falls under the larger domain of multimedia content analysis, which includes the theories, algorithms and systems that extract or infer descriptors which encode characteristics of multimedia content. These descriptors take a variety of forms, ranging from machine interpretable indexing features, to metadata labels in the form of textual words or phrases that can also be  interpreted directly by humans (e.g., ~\cite{Kang:2003, hanjalichighlights:2005, Chan:2005}). The common function of such descriptors is to represent video content in a way that makes possible the systems that give users better access to multimedia content. In particular, here, we are interested in video indexing techniques that will be used for video search engines and other systems that support browsing video collections or otherwise represent to users the contents of a video stream.

Conventionally, video indexing has focused on describing videos in terms of the content that humans identify as being explicitly depicted in their visual channel. Much attention has been devoted to developing algorithms that detect visual concepts in video that are related to events, objects, people, scenes, and locations~\cite{SnoekFNTIR09}. Such concepts can be considered the `literal' content of a video. The meaning or the value of a particular video for a viewer clearly goes far beyond its literal content, however. Videos can also be characterized in terms of how they influence viewers' emotions, i.e., their affective impact on viewers. Affective viewer response refers to the intensity and type of emotion that is evoked in a viewer while watching a video. The potential of affective indexing to contribute to the automatic creation of descriptions that are useful for video search engines is widely acknowledged. However, much research in the area of video indexing remains focused on literal descriptions of video and affective video indexing has yet to reach its full potential. 

An important factor contributing to the success of visual concept detection and other literal approaches to indexing video is the existence of standardized corpora (data sets). These corpora are made available to the research community, often within the framework of a benchmarking initiative, and can be used by researchers to evaluate the algorithms that they develop. For example, detection of visual concepts in video have been a primary focus for the largest multimedia benchmarking efforts, most notably TRECVid~\cite{trecvid:2006, trecvid.features}. Similar large, high-quality data sets, used at the community level in benchmarking initiatives, have yet to be developed for affective video indexing.

This paper takes the position that corpora have a key role to play in supporting the research work that is necessary in order to allow affective video indexing to reach its full potential. The main contribution of this paper is a set of `affective video indexing corpus development guidelines' that arise from a discussion of the state of the art and an analysis of the limitations of existing data sets. The guidelines are organized along three dimensions that are identified as critical for the process of corpus development for affective video indexing:  the context of viewer response, personal variation among viewers, and the effectiveness and efficiency of the process of collecting viewer-reported affective reactions. 
The paper is organized as follows. In the remainder of this section, we set the scene, motivating affective video indexing research and discussing how corpus development contributes to its advancement. 
Section~\ref{sec:emo} and Section~\ref{sec:resources} provide background material on affect in multimedia and discuss existing techniques. 
Then, Section~\ref{sec:affectIndexing} covers previous work on affective video indexing, and the corpora that have been used in that work. Building on the information in Section~\ref{sec:resources}~and~\ref{sec:affectIndexing}, we formulate a set of corpus development guidelines for affective video indexing, which we present in Section~\ref{sec:specs}.  Next, in Section~\ref{sec:ourdatasets}, we introduce three corpora that we have developed using different settings for the collection of viewer affective response: in the laboratory, a Web-based online platform and a crowdsourcing platform. These corpora illustrate progressively more advanced applications of our proposed guidelines. We finish in Section~\ref{sec:conclusion} with conclusions and an outlook on the future of corpus development for affective video indexing.

\subsection{The rise of affective video indexing}
\label{sec:motivation}
The affective video indexing paradigm assumes that users' focus in selecting multimedia content involves a strong affective component and that a multimedia information system, e.g., a video search engine, must be able to take feelings, emotion and mood into account. Recently, the importance of affect in people's information seeking behavior has been recognized, as witnessed by work in the area of conventional text information retrieval, such as~\cite{Arapakis:2008, Lopatovska:2011}. In parallel, awareness of the potential of affective indexing for multimedia information retrieval has also increased. 

Affective video retrieval was first discussed in the mid-1990's by Rosalind Picard as an application of affective computing~\cite{Picard95}. Affective video indexing is well summarized by her statement, ``Although affective annotations, like content  annotations, will not be universal, they will still help reduce time searching for the `right scene'."~\cite{Picard95} (p. 11). When it was first introduced, the proposal that affect could provide an effective means to organize video was not immediately widely accepted. 
%http://hd.media.mit.edu/tech-reports/TR-321-ABSTRACT.html
However, a decade later, the idea had matured in form and established its status as a new paradigm within multimedia information retrieval community~\cite{hanjalic:2006}. 
% Affective content is defined as the `...intensity and type of feeling or emotion (both are referred to as affect) that are expected to arise in the user while watching that clip.'

The importance of affect is now widely accepted by researchers, as reflected by~\cite{Lew:2006}, a survey of multimedia information retrieval, which states that ``On a fundamental level, the notion of user satisfaction is inherently emotional." (p. 3). The current paper is motivated by our conviction that the availability of large, high-quality corpora for the evaluation of affective video indexing will support the multimedia research community in turning its awareness of the importance of affective video indexing into tangible and significant advancement of the state of the art. 

\subsection{The challenge of affective video indexing}
The central challenge faced by affective video indexing lies in the difference between descriptions that refer to the affective impact of videos (e.g., ``uplifting") and descriptions that refer to the literal content of the video (e.g., ``sunrise"). In the case of descriptions of literal content, viewers can quickly and consistently assess whether a description is relevant for a given video, e.g., whether or not a given visual concept is depicted in the video. In making this judgment, they rely on cognitive processing combined with general world knowledge. The judgment is considered to be objective because it can be easily reproduced by consulting a group of viewers, largely independently of the viewers' backgrounds.

Characterizing a video with respect to its affective impact on videos is less clear cut. Information on affect can be gathered by asking viewers to report their emotional response upon watching the video. Affective response is considered to be subjective, since only the subject experiencing the response (i.e., the viewer) is in a position of authority to assess or confirm a particular response. It is tempting to conclude that subjectivity (i.e., the fact that no observer other than the viewer has access to direct knowledge of the viewer's affective response) makes the problem of predicting affective response to a video hopelessly ill defined. Indeed, the affective response evoked in a viewer while watching the video is personal in that it can, and does, differ from person to person. It is also contextual, since it varies when the context in which the video is watched or the underlying mood or physical state of the viewer changes. 

However, although it is not clear cut, affective response is far from arbitrary. In many cases, affective impact will be quite consistent and there will be a high level of agreement in affective response across viewers. The challenge of affective indexing for video involves how to identify those aspects of video that trigger emotional reactions across viewers that are stable enough that they can be robustly predicted.

The stability of affective impact is most clearly illustrated in the case of film. Filmmakers are highly skilled in evoking specific emotions in their audience. The high-level of inter-subjective agreement concerning the connotative aspects of film has been studied and used as the basis for an automatic indexing system by~\cite{Benini:2011}. Connotation is that dimension of interpretation that goes beyond literal meaning, and, as such, encompasses a large affective component.  Today's video search engines index large quantities of video on the Web. For online videos, it is not possible to apply the a priori assumptions about the the techniques and conventions used in formal film to trigger emotions. However, it is possible to anticipate that there will be a component of viewer response that is grounded in modalities of emotional reaction shared in the audience or arising from common interpretation conventions. 

This paper takes the standpoint that by isolating and emphasizing aspects of video for which human judges display a relatively high level of agreement, corpora for the evaluation of affective indexing can be created that can make a contribution to advancing the state of the art comparable to the contribution made by benchmarks that focus on literal descriptions of video content.

\subsection{The contribution of corpus development}
Corpus development contributes to advancing the state of the art of multimedia technology by making possible standardized evaluation. Only when a standard data set and ground truth are used, is it possible to directly and fairly compare alternative algorithms. Comparison and reproducibility help to drive forward the state of the art: when researchers know how their algorithms perform with respect to the state of the art, they can better direct their efforts to surpass it and more quickly abandon less promising lines of investigation.  Benchmarks and standard tasks/data sets help to eliminate redundancy by enabling direct comparison between algorithms across research sites, increasing the efficiency of the research community by allowing resources to be shared between sites and providing a framework in which researchers can interact in a mixture of collaboration and competition that is stimulating and productive. The impact of the TRECVid evaluation for video has been large and is well documented~\cite{Thornley:2011}. However, as mentioned above, TRECVid focuses on literal approaches to video indexing, i.e., content explicitly depicted in the visual channel. Corpus development is key to allowing affective video indexing to achieve similar impact.

Corpus development does not strive to promote one particular variety of affective video indexing, but rather if numerous, well-designed multimedia corpora were available, they would contribute in many different ways. Here, we provide some examples of the range of applications in which affective video indexing, retrieval and browsing has been used. In~\cite{rui:2000}, highlights were extracted from baseball programs and in~\cite{hanjalichighlights:2005} an adaptive approach to sports video highlight detection was proposed and studied in detail for the case of soccer. The usefulness of such affective video indexing techniques is witnessed by the fact that Mitsubishi has already released two products taking advantage of the highlight detection for sport events in Japan \cite{mitsu-dvd}. Retrieval of movie clips using multimedia content features and user-assigned keywords was investigated by~\cite{Chan:2005} and~\cite{jones:2011}. Laughter events have been successfully deployed in videos for navigation~\cite{Janin:2010}. 
%Then we have `Modeling affective evaluation of multimedia contents: user models
%	to Associate subjective experience, physiological expression and
%	contents description.'~\cite{EURECOM2361}.
%Does it actually make sense that these citations are mutually exclusive from what we have in the table?
The examples illustrate the spread of application areas that stand to benefit if large, high-quality corpora can be developed in made available to the research community.

%The time is ripe for this paper because:
%\begin{itemize}
%\item Cognitive concept detectors are revealing their limitations.
%\item We have access to large numbers of people that can do annotations.
%\item The amount of video is so huge that even if we had perfect cognitive concept detectors, we would need to have some other way of choosing video. Affective response is a natural alternative, since it is known to play an important part in our decision making processes.
%\item Existing corpora have been limited in scope and applicability. As a community, we should make a conscious effort to counter this trend.
%\end{itemize}

%This paper brings together background information and examples of affective video indexing data sets with the aim of supporting researchers in developing data sets for affective video indexing. 

Current data sets used to evaluate individual theories and algorithms in affective content analysis are typically limited in size and scope. The limitations are imposed because of the relative difficulty of collecting affective responses from many viewers. These limitations also reduce variability in the elicited affective responses of test users, which facilitates manual annotation and results interpretation, but may ultimately be too narrow for the resulting algorithms to be used in practical situations. 
%Shortcomings of currently available data sets

Recently, however, technological developments have provided means for developing a new generation of corpora.
%The Web now offers new opportunities for creating the data sets necessary to develop and evaluate algorithms for affective indexing of video. 
Online systems make it possible to ask large numbers of viewers to watch videos and provide information on their affective response. Additionally, the rise of crowdsourcing and large crowdsourcing platforms such as Amazon Mechanical Turk (www.mturk.com) make it possible to more easily recruit large numbers of annotators with a representative spread of backgrounds. Corpora in existence today do not, as yet, fully exploit these resources.  The corpus development guidelines set out in this paper aim to encourage the effective use of the new opportunities offered by the Web and by crowdsourcing platforms.

In this study, we set our focus on corpora used to study emotion evoked by video for the purpose of affective video indexing.
It is important to note that video corpora are also developed in order to study human emotion directly.
Such corpora are developed for the general goal of studying emotion.
For example, videos of people laughing would be used to study expressions of positive emotion.
Research involving such corpora has been treated elsewhere in the literature~\cite{4468714, morency2011towards, Biel2013}. 
Here, we are interested in the emotion of people watching the video, which is not necessarily the same as the emotion of people appearing in the video.
For example, videos of people laughing can either evoke a positive or negative emotion in viewers, depending on their perceptions of the person who is laughing.

A concise discussion of the importance of data sets for affect recognition technologies, emphasizing the pressing need for new corpora of emotion-eliciting films, is included in~\cite{morris2011crowdsourcing}. In order to cover the full scope of affective computing, multimedia corpora designed to elicit the complete spectrum of human emotional response are necessary. Such corpora should not, a priori, be assumed to be useful for the purposes that we address here, namely, affective video indexing. As discussed further below, the types and distributions of emotion triggers found in video may be different than those occurring in more general contexts. However, the message of~\cite{morris2011crowdsourcing}, that large amounts of data must be collected in order to carry out research in the area of affect, is a key point that is common with our own work. Further, like our work,~\cite{morris2011crowdsourcing} identifies crowdsourcing as providing a highly promising new opportunity to gather emotional response data from human subjects.

\vspace{-9pt}
\section{Emotion in response to multimedia}
\label{sec:emo}
In this section, we provide a specification of the key concepts of affect and multimedia that are used in this paper and cover the relevant related work. We start out by presenting a clear definition of affective viewer response to multimedia as it is applied for affective video indexing and we continue to discuss this definition with respect to the larger field of research on human emotion.
\vspace{-19pt}
\subsection{Affective viewer response in context of affective video indexing}
\label{sec:aviresponse}
Affective viewer response, as already mentioned above, is the intensity and type of emotion that is evoked in a viewer while watching a video. 
Emotions are complex phenomena with affective, cognitive, conative and physiological components~\cite{citeulike:8222108}. The affective component is the subjective experience conventionally connected with feelings. The cognitive component is the perception and evaluation of the emotional situation. The conative component is the expression of affect, including facial expressions, body gesture, and any other action that has a preparatory function for an emotional situation. The physiological component regulates physiological responses in reaction to the emotional situation, for example, increasing perspiration during a fearful experience. When studying emotion evoked in viewers in response to multimedia, it is important to take the complexity of emotion into account rather than expecting emotion to manifest itself along a single dimension only.

The factor that differentiates the study of affective viewer response to multimedia for the purpose of affective video indexing from other studies of human emotion is its focus on the trigger of emotions. While many researchers strive to understand human emotions, researchers in the area of affective video indexing have the objective of describing the content of the video that \emph{causes} these emotions. Consequently, affective video indexing researchers concentrate on emotions felt by viewers in response to video, and not on the entire spectrum of possible emotions. Although it is not possible to completely exclude that a viewer might experience any specific emotion while viewing a video, in general, we note that the type and particular the frequency of emotion evoked by video should not be assumed to be the same as those in other situations. This point can be highlighted by reference to the work of~\cite{riek2011guess}, who study a set of YouTube videos portraying moods reported to be consistently happy. If happiness is indeed frequently represented in videos on YouTube, researchers in affective video indexing need to take into consideration that portrayed happiness may possibly evoke a specific affective viewer response more frequently than it would be evoked in other settings. Recall that the emotional responses of viewers to happiness portrayed in a video is still expected to cover a wide range, potentially including emotions generally considered positive, as well as those generally considered negative.

Clearly, it is not useful to draw an overly sharp delineation between the study of viewer response for the purpose of affective video indexing and for other purposes. Some affective video indexing corpora will contain selected video content that allows researchers to focus specifically on video content that triggers particular emotional responses. However, careful attention to the difference between corpora for affective video indexing and other video corpora used to study emotions from different perspectives helps to ensure the suitability of the affective corpus for its particular purpose. In the next subsection, we move on to a more general discussion of research on human emotion, maintaining emphasis on that aspects particularly important for affective viewer response to video.
\subsection{Research on human emotion}
Because of the complexity of emotion, a good entry point into research on human emotion is a clear picture of what emotion is \emph{not}. In particular, in understanding the nature of emotional response, the terms ``mood" and ``emotion" should be differentiated. The point is a particularly important one, since these terms are sometimes used interchangeably in the literature despite the clear formal distinction between their definitions. Mood is a diffused affective state that is long, slow moving and not tied to a specific object or elicitor whereas emotions can occur in short moments with higher intensities~\cite{scherer2005}. 
%Scherer characterizes the distinction using the notions of intrinsic and extrinsic appraisal~\cite{scherer2005}. Intrinsic appraisal is independent of the current goals and values of the viewer while extrinsic or transactional appraisal leads to feeling emotions in response to the stimuli. 
%For example, the intrinsic emotion of an image depicting someone who is smiling is happiness. If the person smiling is a figure disliked by the viewer, extrinsic appraisal leads to unpleasant emotions. In this paper, we are concerned with extrinsic emotion, with video as the elicitor of the emotional response.
 
One of the most well-known and widely-accepted theories that explains the development of emotional experience is appraisal theory. According to this theory, cognitive judgment about, or appraisal of, a situation is a key factor in the emergence of emotions~\cite{citeulike:2821039,ortony88emotion,citeulike:3014462}. According to Orthony, Clore and Collins (OCC)~\cite{ortony88emotion}, emotions are experienced following a scenario comprising a series of phases. First, there is a perception of an event, object or an action. Then, there is an evaluation of the event, object or action according to personal wishes or norms. Finally, perception and evaluation result in a specific emotion or emotions arising. Considering this scenario for an emotional experience in response to multimedia content, emotions arise first through sympathy with the emotions that are depicted in the content~\cite{citeulike:8222108}. 
%can it also be the presented situation?
During the appraisal process for an emotional experience in response to multimedia content, viewers examine events, situations and objects with respect to novelty, pleasantness, goal, attainability, copability, and compatibility with their norms. Then, the viewers' perceptions induce specific emotions, which changes their physiological responses, motor actions, and feelings. 

Emotional processes can be divided into different categories. Here, we mention three processes that apply not only in the general case of emotional response, but also in the specific case of viewer affective response to video: \textit{emotion induction}, \textit{emotional contagion} and \textit{empathic sympathy}~\cite{citeulike:8222108}. An example of \textit{emotion induction} is when in a TV show a politician's comment makes the viewers angry while the politician himself is not angry. The angry response from the viewers is due to their perception of the situation according to their goals and values. \textit{Emotional contagion} occurs when the viewer only experiences the emotion expressed in the video. For example, the induced joy as a result of sitcom laughter falls into this category. In the empathic category, the situation or event does not affect the viewer directly, but rather the viewer reproduces the appraisal steps of the characters who are depicted in the video. The empathic reaction is  described as \emph{symmetric co-emotion} in cases in which the viewer has positive feelings about the character and \emph{asymmetric co-emotion} in cases in which the viewer has negative feeling about the character~\cite{citeulike:8227534}.

Empathy is a complex phenomenon with both cognitive and affective components. Affective empathy is the primitive response involved in sympathizing with other individuals. On the other hand, cognitive empathy is the intellectual understanding of other people and the rational reconstruction of their feelings~\cite{citeulike:8227551,citeulike:8222108}. Zillman developed an affective disposition theory for narrative plot~\cite{citeulike:8227621,citeulike:8227534}. According to this theory, empathic emotions originate with the observation of the actors by viewers. First, a character's actions are morally judged by the viewer and the judgment results in a positive or negative perception of the character. Then, depending on whether the viewer approves or disapproves of the character, the viewer sympathizes either empathically or counter-empathetically. The intensity of the affective response to a video depends on how much viewers identify themselves with the protagonists and to what extent they suspend their own identities~\cite{citeulike:8227621}.

In general, it is challenging to go from appraisal theory and the sources of affective empathy to a technique that analyzes the content to predict viewers' emotions. However, some characteristics of video are quite indicative of affective response. Much video will capture not only action, but the audience of that action. This audience might be the spectators of a sports event or certain characters in a film, who watch and react to events. The reaction of these in-video observers (e.g., laughter or cheering) can provide important clues to how viewers will react to the video. Further, the literature has identified the most important emotion inducing components of movies as being music and narrative structures~\cite{citeulike:8222108}. Music is clearly instantiated at the signal level, and structure can also to a certain extent be extracted (e.g., the quick shot changes of a chase scene that will correspond to abrupt changes in the visual constitution of a scene). In pursuit of such regularities, researchers have undertaken to develop techniques for affective video content analysis, which we will return to discuss in Section~\ref{sec:affectIndexing}.
%ML: 18/11/2011: The "Affect in Multimedia Search" section was split between the introduction and the goals of benchmarking section
%\section{Affect in Multimedia Search}
%In this section, we first discuss in more detail the difference between video indexing that is focused on cognitive aspects and video indexing that includes affective aspects. We then overview the literature on multimedia information retrieval that applies affective indexing. The picture that emerges is that affective video indexing has a enormous potential and stands to benefit from strengthening and focusing of research efforts via the development of corpora and tasks for affective video indexing.
%\label{sec:affectinMIR}

\section{Background and Existing Techniques}
\label{sec:resources}
%\textbf{Note4MS: We will need a new introduction to this section describing its content and function}
This section discuses emotional representations and existing tools that have been developed to collect annotations in the form of these representations are introduced.

\subsection{Emotional representations}
There are different emotional representations including, discrete, and continuous models. 
Discrete representations of emotions, and their theoretical underpinnings, were originally inspired by the representation scheme of Darwin, who considered emotion important for survival. 
%Mohammad: here I changed it so it is clearly about humans only: please check.
Discrete representations presuppose the existence of the certain number of basic and universal emotions~\cite{scherer2005,citeulike:8845560}. Some of the most widely-known research on basic emotions was carried out by Ekman~\cite{citeulike:8845560}, whose work supported the universality of facial expressions of emotion.
There is currently no answer to the question of how many different emotions there are, but most lists in use contain 6-14 emotions~\cite{scherer2005}.
The lists of emotions used by psychologists are generally utilitarian in meaning. Utilitarian emotions are those that are helpful for adapting to the world and make an important contribution to our well-being~\cite{scherer2005}.
%The basic emotions are mostly utilitarian emotion. 
%and their number is usually from 2 to 14. 
%Scherer suggests using the term ``modal'' instead of ``basic'' emotions. 
%Mohammad: could you please add a sentence that explains *why* he uses the term "modal", or at least why we use both terminologies in the paper instead of sticking to either one of the other.
It is important to note that utilitarian emotions may not be the most important emotions for the purpose of affective video indexing.
Another type of emotion represented with discrete categories is aesthetic emotions~\cite{scherer2005}, emotions arising in conjunction with the appreciation of beauty or quality (including, `admiration', `ecstasy' and `fascination'.) 

When human affective response is conceptualized as a discrete set of categories, particular challenge arises. The challenge involves ensuring that the categories are interpreted in the similarly across different situations. A readily observable difficulty in explicitly defining the scope and coverage of individual affective categories is cross-lingual variation. Such variation arises since emotion words often do not have exact translations into different languages, e.g., there is no word in Polish that corresponds exactly in meaning to the English word, ``disgust''~\cite{Russell1991426}. Scherer~\cite{scherer2005} takes a pragmatic step towards addressing this issue by proposing a mapping between words or word stems, as used by subjects in free choice emotional reports, to a set of 36 affect categories. Being limited to the English language, this mapping does not, of course, address the cross-lingual issue, but it makes an important contribution to the comparability of emotion reports and emotion studies. 

The challenge of ensuring that emotion categories receive a consistent interpretation contributed to the motivation for the development of dimensional approaches.
Wundt~\cite{citeulike:9352253} was the first to propose a dimensional representation for emotions. Recent tools that have been developed, e.g., by~\cite{citeulike:2239554}, discussed in more detail below, %are more recent reflexes of the effort 
 use dimensional approaches to minimize the effect of differences in interpretations of discrete categories and to verify inter-participant consistency, e.g., Bradley and Lang \cite{citeulike:2239554}. 
Dimensional theories of emotion are grounded in the that emotions can be represented in a continuous space and that discrete emotions are basically folk-psychological concepts that can be identified with points in this space~\cite{citeulike:9343328}. 

Dimensional representations used by psychologists often represent emotions in an n-dimensional space (generally 2- or 3-dimensional). 
The most well-known example of such a space arises is the 3D valence-arousal-dominance or Pleasure-Arousal-Dominance (PAD) space~\cite{citeulike:2642710}. 
%Mohammad: Can we add here what "PAD" stands for? It is not clear from the text.
This space arises from cognitive theory and is widely used for studying affect and multimedia---we ourselves make use of it for corpus development, as discussed later in the paper. 
The valence scale ranges from unpleasant to pleasant. The arousal scale ranges from passive to active or excited. The dominance scale ranges from submissive (or ``without control'') to dominant (or ``in control, empowered''). 
Fontaine et al.~\cite{Fontaine01122007} proposed adding predictability dimension to PAD dimensions. Predictability level describes to what extent the sequence of events is predictable or surprising for a person. 

An advantage of using dimensional representations of emotion is that when people are asked to describe their emotions, they are often better at positioning content in comparison to a reference point, i.e., this video was more exciting than the previous one, compared to the situation where they are asked to provide an absolute score~\cite{4959919}. Methods related to those used by~\cite{EURECOM2361} are good examples of approaches that make use of relative annotation.

It is important to note that dimensional representations of emotion should not be considered to supersede discrete representations. 
Rather, both have an important contribution to make to facilitating consistent representation of emotional response in the face of variability introduced by different subjects, different triggers, different contexts and also the passage of time.
The importance of the way in which people talk about their own emotions should not be underestimated.
Although psychologists strive to develop representation systems transcending folk-psychological concepts, human language remains an important tool for gaining insight on human emotion.
The robustness, reproducibility and relevance of natural language characterizations of emotions is assured via the very process which gave rise to the development of human language to support inter-human communication.
The critical factor guiding the development of human language expressing emotion can be considered to be the need for the effective and reliable communication of emotions between humans within the context of a shared understanding of a common emotional space.
Recent work has been devoted the creation of computational models of the conceptual meaning of words that humans used to describe emotion~\cite{6496208}. The models are able to map between emotion vocabularies of different sizes in different languages and also to differentiate between nuances of meaning in large emotion word vocabularies.

A particularly important consideration in the area of affective video indexing is that the emotional representation used must provide a good fit with the types of emotion that are expected to occur in response to triggers in video content. In other words, in order to understand the emotional categories or the dimensions that are most appropriate for assessing viewers' emotion, the video content and the context in which it is being watched must be carefully considered.  Dimensions can be also identified based on the application. For example, the arousal of a  viewer while watching a live sports event might be substantially different than the arousal of a viewer watching a replay of the same sports match. 

Finally, in gathering emotional annotations, co-occurring emotions should be also considered. There are cases of co-occurring emotions from different poles of the spectrum or an emotion that co-occurs with a feeling of ambivalence. For example, some music listeners when asked to report their emotions for positive and negative affect separately, rated both as high \cite{hunter2008mixed}. This might be a result of music listeners' preference for sad stimuli due to their underlying mood~\cite{hunter2011misery}.

%Mohammad writes this
\subsection{Emotional self-reporting methods}
%Mohammad: check this first sentence, I added the word "underlying" because I think that "true" sounded a bit vague.
Understanding the ``true", underlying emotion that was felt by a participant during an experiment has been always a challenge for psychologists. Multiple emotional self-reporting methods have been created and used so far \cite{Desmet2003,Winoto20106086,scherer2005,citeulike:2239554,citeulike:2772174}. %However, none of them give a generalized, simple and accurate means for emotional self-reporting. 
Emotional self-reporting can be done either in free-response or forced-choice formats. In the free-response format, the experiment participants are free to express their emotions by words. In the forced-choice, participants are asked to answer specific questions to indicate their emotions. Forced-choice self-reports in affective experiments use both discrete and dimensional approaches. Discrete-emotion self-reporting tools have been developed that can be used to ask participants to report their emotions with emotional words on nominal and ordinal scales. Dimensional approaches of emotional self-reporting are based on bipolar dimensions of emotions. Emotions can be reported along each dimension using ordinal or continuous scales~\cite{citeulike:2239554}. Here, we discuss in more detail some popular self-reporting methods which have been used for psychological and human computer interaction research.

Russell \cite{Russell1980} introduced the ``circumplex model" of affect for emotion representation. In his model, eight emotions; namely, ``arousal", ``excitement", ``pleasure", ``contentment", ``sleepiness", ``depression", ``misery" and ``distress" are positioned on a circle surrounding a two dimensional activation, pleasure-displeasure space. Starting form these eight categories, 28 emotional keywords were positioned on this circumplex, based on the results of a user study. The advantage of this circumplex over either discrete or dimensional models is that all the emotions can be mapped on the circumplex using only the angle. In this way, all emotions are presented on a circular and one dimensional model. 

\begin{figure}[ht]
\begin{center}
\includegraphics[width=0.5\linewidth]{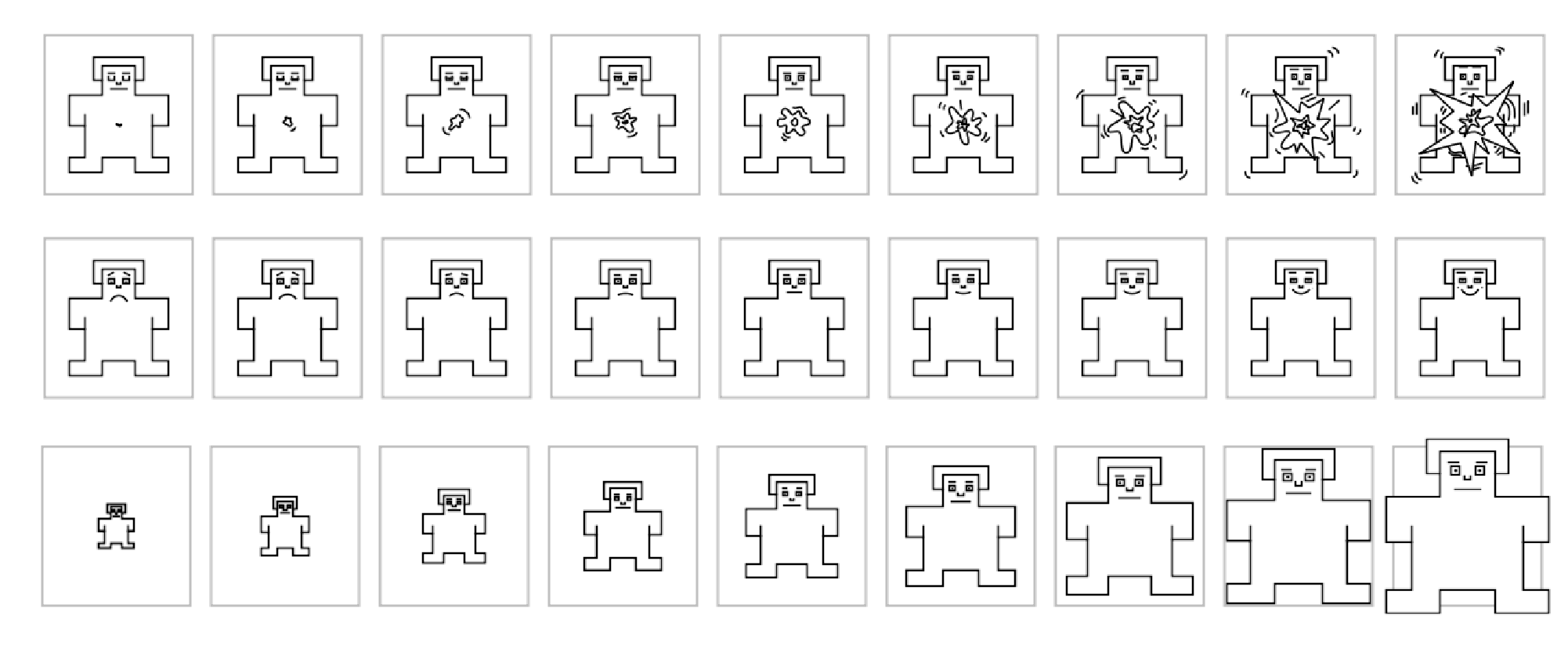}
\caption{Self Assessment Manikins. From top to bottom the manikins express different levels of arousal, valence, and dominance.}
\label{fig:sam}
\end{center}

\end{figure}

The Self Assessment Manikin (SAM) is one of the most well-known emotional self-reporting tools. It consists of manikins expressing emotions. The emotions vary along three different dimensions; namely, arousal, valence, and dominance \cite{citeulike:2239554}. The SAM Manikins are shown in Fig. \ref{fig:sam}. Experiment participants can choose the manikin that best portrays their emotion. This method does not require the verbalization of emotions and the manikins are understandable without further explanation. For these reasons, the SAM tool is language independent. The second advantage of the SAM tool is that it can be directly used in measuring the multiple dimensions of emotions. A limitation of SAM is that subjects are unable to express co-occurring emotions with this tool. 

The ``Positive and Negative Schedule" (PANAS)~\cite{citeulike:8384089} permits self-reporting 10 positive and 10 negative affects on a five-point scale. An expanded version of PANAS, the ``Positive and Negative Schedule---Expanded Form" (PANAS-X), was developed later.  PANAS-X provides the possibility of reporting 11 discrete emotion groups on a five-point scale~\cite{citeulike:8758783}. PANAS is made to report affective states and can be used to report both moods and emotions. PANAS-X includes 60 emotional words and takes on average 10 minutes for an experimental participant to complete~\cite{citeulike:8758783}. The time needed to answer the PANAS questionnaire makes it too difficult to use in the experiments with limited time and multiple stimuli.

\begin{figure}[ht]
\begin{center}
\includegraphics[width=0.5\linewidth]{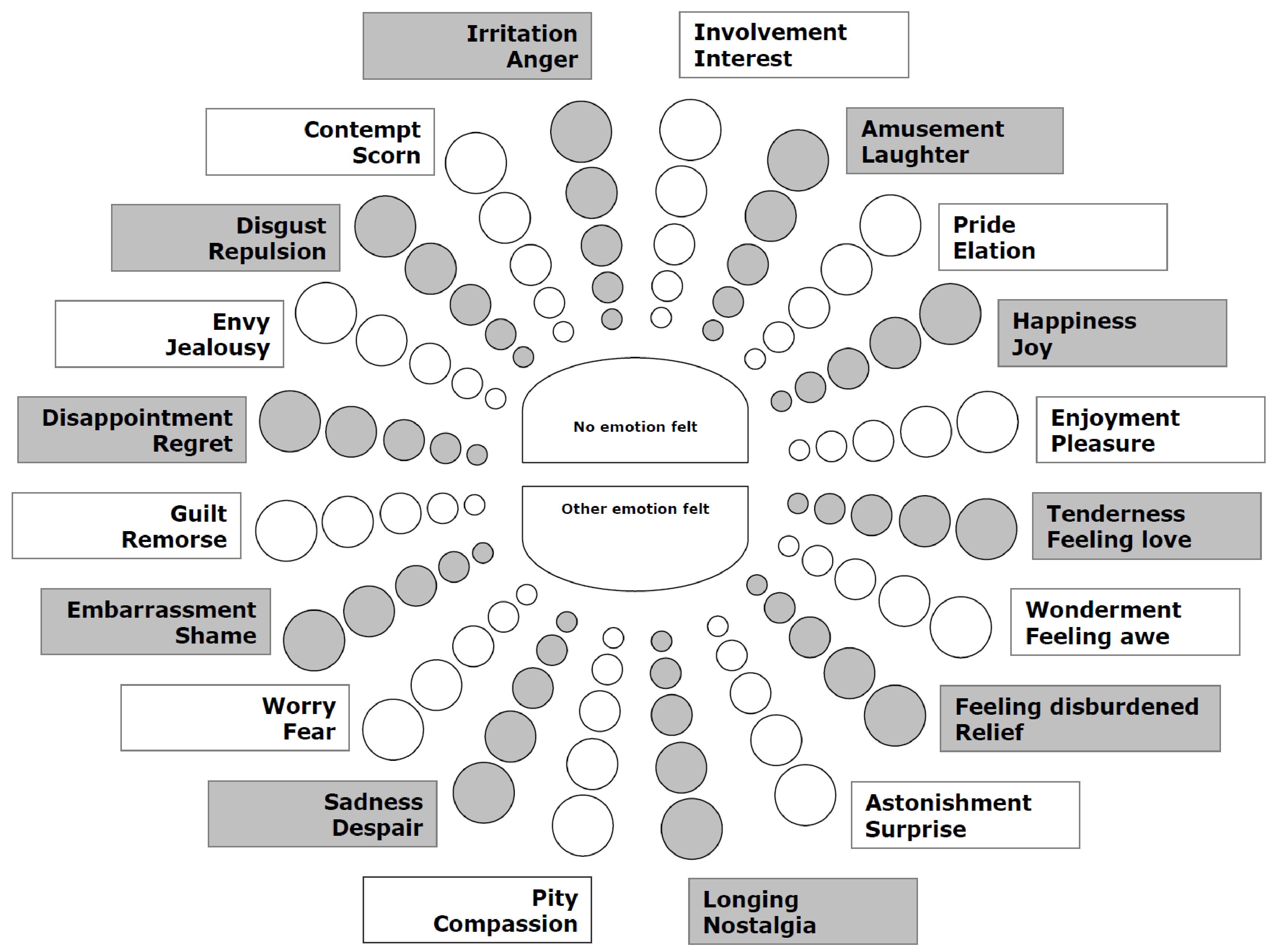}
%\caption{A participant can indicate his emotion on Geneva emotion wheel by clicking or choosing the circles.}
\caption{Subjects can indicate their emotion on the Geneva Emotion Wheel~\cite{scherer2005} by selecting the corresponding circle.}
\label{fig:gvEmotionWheel}
\end{center}

\end{figure}

Scherer~\cite{scherer2005} positioned 20 emotions around a circle to combine both dimensional and discrete emotional approaches, and in this way created the Geneva Emotion Wheel. For each emotion around the wheel, five circles whose size increases from the center outwards are displayed. The size of the circles is an indicator of the intensity of felt emotion (see Fig.~\ref{fig:gvEmotionWheel}). In an experiment, participants can pick, from the list of 20 emotions, up to two emotions that were the closest to their experience and report the intensities of the emotions with the size of the marked circles. In case no emotion is felt, a user can mark the upper half circle in the hub of the wheel. If a different emotion is felt by a user, it can be indicated in the lower half circle. The emotions are sorted on the circle such that, high-control emotions are on the top and low-control emotions are at the bottom and the horizontal axis, which is not explicitly visible on the wheel, represents valence or pleasantness.

%Mohammad: Can you add a sentence that makes explicit how PrEmo deals with co-occurring emotions? For me, it's not completely obvious from reading this description.
PrEmo is an alternative non-verbal emotion reporting tool to report emotions in response to product design~\cite{Desmet2003}. It overcomes the problem of reporting co-occurring emotions by making use of animated characters expressing emotions. PrEmo consists of 14 animated characters expressing different emotions and it is, for this reason, language independent. Users can assign a score, at three levels, to one or more characters that they identify as relevant to their emotional response.
%
%\begin{figure}[ht]
%\begin{center}
%\includegraphics[width=0.5\linewidth]{figs/premointerface.eps}
%\caption{Users can identify emotion they are feeling with 14 animated characters are expressing emotions.}
%\label{fig:prEmo}
%\end{center}
%\end{figure} 
%\vspace{-1.5em}
%Zoghbi et al. proposed a joystick for  emotional communication in human robot interactions \cite{Zoghbi:2009:EAS:1732643.1732664}. Users can report the pleasantness or valence by moving the joystick forward and backward and report the level of arousal by squeezing the joystick handle. Although this tool gives the possibility to report the emotions in real time it causes distraction for a participant in front of a stimulus and it is not yet available for other researchers to use.

\subsection{Video affective annotation tools}

%Tools that are available (feeltrace, Mohammad's annotation tool) Eurecom feeltrace-type tool (software)
%drag and rop multimedia d'n'dMultimedia 

Among existing self-reporting tools, few have been designed specifically for the affective annotation of video. Villon developed an annotation tool with which a user can drag and drop videos onto the valence-arousal plane~\cite{EURECOM2361}. This tool presents the possibility of comparing the ratings  given to different videos and enables an experiment participant to rate a video relative to the ratings given other videos. The tool enables users to take their previous reports into account while annotating a new video.

Feeltrace was developed to annotate the intrinsic emotion in videos~\cite{citeulike:3721917}. 
%Mohammad: We need to define intrinsic emotion...maybe even in the "Emotion in response to multimedia" subsection of Section 2.
This tool was originally designed to annotate the emotions that are expressed by people who are depicted in videos (e.g., in talk shows), including acted facial expressions or gestures~\cite{citeulike:7921301}. Although this tool provides the possibility of continuous annotation, it is not a particularly appropriate tool for emotional self-reporting, because it is difficult for tool users to both concentrate on the video and at the same time changes in their own emotions. 

An online video affective annotation tool has been developed by Soleymani et al.~\cite{5349526}. With their annotation tool, a experiment participant can self-report emotions after watching a given video clip by means of SAM manikins and emotional keywords from a selected list in a drop down menu. This tool is used in the development of our web-based corpus presented in Section \ref{subsec:webBased}. % (see Fig. ~ref{fig:snapshot}). 

%Mohammad writes this
%A. Existing corpora (listing existing corpus and discussing them with respect to the three objectives for corpus %development above)
%1. Semaine
%2. Rottenberg
%3. etc.

%B. Our examples
%The practical examples include definition of the task (i.e., what the corpus is designed to allow us to achieve), description of the method of collection and analysis of collected annotations.
%1. Lab-based (Mohammad's work)

\setlength{\tabcolsep}{5pt}
\begin{table*}[h!t]
\renewcommand{\arraystretch}{1.3}
%Mohammad: Can you check this?
%\caption{The summary of video affective representation literature.}
\caption{Key examples of previous work on multimedia content analysis for affective video indexing and existing corpora.}

\begin{center}
\begin{tabular}{p{0.6cm}|p{1.8cm}|p{6.9cm}|p{1.4cm}|p{1.4cm}|p{3.7cm}}
\hline\hline
\noalign{\smallskip}
\label{tab:litReview}
\textbf{Study} & \textbf{Emotion repr.}& \textbf{Categories or dimensions}& \textbf{Nr of \newline Annotators} & \textbf{Modalities} & \textbf{Evaluation results}\\\hline
\cite{Kang:2003} & disc. &fear/anger, joy, sadness and neutral& 10& V & classification rate, fear: 81.3\%, sadness: 76.5\%, joy: 78.4\%\\\hline
\cite{hanjalicxu:2005} & cont. & valence and arousal & - &  AV& no evaluation\\\hline
\cite{1637510} & disc.& fear, anger, surprise, sadness, joy, disgust and neutral & 3 & AV & 74.7\%\\\hline
\cite{4674668} & cont. & pleasure, arousal, and dominance & 14 & AV &-\\\hline
\cite{Xu:2008:HMA:1459359.1459457} & disc. & fear, anger, happiness, sadness and neutral&unknown & AV& 80.7\%\\\hline
\cite{citeulike:7927300} & disc. \& cont.& continuous arousal for shots, positive/negative excited \& calm on scene level& 1&  AV \& text & 63.9\%\\\hline
\cite{5571819} &disc. &  acceptance, anger, anticipation, disgust, joy, fear, sadness, surprise and netural &16& AV &subject agreement rate 0.56\\\hline
\cite{citeulike:8161572} & cont.&valence and arousal & 10&AV& -\\\hline
\cite{citeulike:8638639} & disc. & fear, anger, surprise, sadness, joy, disgust& 6&AV& 71.4\%\\\hline
\cite{DBLP:conf/eccv/DemartyPGS12} & violence& violent \& non-violent & 7& AV & -\\\hline
\end{tabular}
\end{center}

\end{table*}

\section{Affective Video Indexing}\label{sec:affectIndexing}
In this section, we provide an overview of affective indexing and discuss the corpora that have been developed in previous work. In particular, we discuss the  shortcomings of existing corpora, which are used as a basis to develop a set of guidelines for the design and development of future corpora.

\subsection{Affective video analysis for Indexing}

Affective video content analysis involves estimating the affective response elicited in viewers by the content. Motivated by work in the area of film, researchers have extracted content features, such as audio energy and color histograms from the video signal and used machine learning techniques to infer which emotion would be felt by an average viewer. They have considered different goals and applications for their algorithms, from video summarization to personalized content delivery. 

A summary of key examples from the existing literature in the area of content analysis for the emotional understanding of videos is given in Table \ref{tab:litReview}. The table contains information about the data set that was used to evaluate the proposed algorithms, including information about the type of representation used (discrete vs. continuous), the affective categories used, the number of human annotators used to create the corpus, the modalities contained in the video data set and finally, the results of evaluation, if they are given by the paper.

Existing approaches for content analysis generally make use of low-level content features extracted from both video and audio. Low-level audio features often include short time energy, zero crossing rate, Mel Frequency Cepstral Coefficients (MFCC), and pitch. Low level visual features include, color variance, motion component, shot change rate, key lighting, brightness, and color energy~\cite{Xu:2008:HMA:1459359.1459457, Kang:2003,hanjalicxu:2005,4674668}. Irie et al.~\cite{5571819} proposed using a bag of audio-visual words strategy to transform the feature space before classification. 

Different machine learning models have been used to classify videos on different levels, e.g., shots, scenes, into different emotional classes, with the goal of emotional tagging. Kang~\cite{Kang:2003} used a Hidden Markov Models (HMM) classifier to detect emotional events from low level features. Hanjalic and L.-Q. Xu~\cite{hanjalicxu:2005} applied a regression model to predict arousal and valence along a continuous temporal dimension. M. Xu et al.~\cite{Xu:2008:HMA:1459359.1459457} proposed using a hierarchical approach that first clusters the samples in the arousal dimension and then classified them using HMMs into valence classes. Soleymani et al.~\cite{citeulike:7927300} used movie genres and the temporal dimension to estimate emotions elicited by movie scenes using a Bayesian framework. Irie et al.~\cite{5571819} used a latent topic model by defining affective audio-visual words in the content of movies to detect emotions in movie scenes.  This model takes into account temporal information, i.e., the effect of the emotion from the preceding scene, to improve affect classification. The probability of emotional changes between consecutive scenes was also used in~\cite{citeulike:7927300} to improve emotional classification of movie scenes using content features. 

Going beyond multimedia content analysis, emotional responses of the viewers have been also used to detect affective tags for videos. Joho et al.  \cite{citeulike:8161572,Joho:2009:EFE:1646396} developed a video summarization tool using facial expressions. Kierkels et al. \cite{citeulike:6922056} proposed a method for personalized affective tagging of multimedia using peripheral physiological signals. Valence and arousal levels of participants' emotion when watching videos were computed from physiological responses using linear regression \cite{Soleymani:2009:IJSC}. 

%Emotional characteristics of videos have also improved music and image recommendation. 
%Shan et al. \cite{Shan2009music-recom} used affective characterization using content analysis to improve film music recommendation. 
%Tkal\v{c}i\v{c} et al. showed how affective information can improve image recommendation \cite{citeulike:7965793}.  %In their image recommendation scenario, affective scores of images from the international affective picture system (IAPS) \cite{IAPS2005} were used as features for an image recommender. %They conducted an experiment with 52 participants to study the effect of using affective scores. The image recommender using affective scores showed a significant improvement in the performance of their image recommendation system. 

%which were audio energy and visual change rate from videos to create an affective curve in the same way as the affective highlighting method proposed by Hanjalic \cite{hanjalichighlights:2005} 

We next survey existing corpora and point out their strengths as well as their shortcomings that necessitate the development of new corpora. 

\subsection{Existing corpora}
\label{sec:examples}

In this section, we provide a summary of corpora that have previously been developed and used for evaluating affective video indexing. In general, affective video corpora are developed with specific goals. Three common goals are: first, emotion elicitation or mood regulation in psychological experiments; second, emotional characterization of videos using content for video indexing or highlighting and third, recognition of the intrinsic emotions in the videos, e.g., detecting the emotions which were expressed by people in the videos \cite{4468714, mahnob_HCI}. Although all three involve emotion, it is critical to avoid mixing these three different research tracks and the goals behind them. For example, movie excerpts that  are most likely to elicit strong emotions are chosen for emotion elicitation in the case of the first goal. In contrast, for the second goal, an exclusive focus on strongly emotional excerpts is not appropriate for emotional characterization. Emotional characterization should be able to deal with the full spectrum of emotions in videos, from neutral videos to mixed and strong emotions.

Rottenberg et al.~\cite{citeulike:8017077} created an emotional video data set for psychological emotion elicitation studies. The excerpts, which were about 1--10 minutes long, were either extracted from famous commercial movies or from non-commercial videos that were used in emotional research, e.g., an amputation surgery video. First, they formed a set of excerpts with different targeted emotions; namely, amusement, anger, disgust, fear, neutral, sadness and surprise. 
%Mohammad: Can you check this? I think too much is commented out here.
They evaluated the excerpts based on ``intensity'' and ``discreteness''. The ``intensity'' of an excerpt means whether a video received high mean report on the target emotion in comparison to other videos. The ``discreteness'' refers to the extent to which the target emotion was felt more intensely in comparison to all non-targeted emotions.  ``Discreteness" was calculated using the ratings a video received on the target emotion in comparison to the other emotions. Ultimately, the data set that was formed consisted of 13 videos, from under a minute to up to eight minutes long, for emotion elicitation studies.

In a more recent study, Schaefer et al.~\cite{citeulike:8384089} created a larger data set from movie excerpts to induce emotions. The study went beyond discrete basic emotions and developed a corpus including 15 mixed feelings in addition to six discrete emotions; namely, ``anger", ``disgust", ``sadness", ``fear", ``amusement", and ``tenderness". 364 participants annotated their database using three questionnaires. %After watching each video, participants answered emotional arousal on seven points scale. Then using a modified version of Differential Emotions Scale (DES) questionnaire, they reported how much they felt each of the 16 listed emotions on a seven point scale. The third questionnaire was PANAS with 10 positive and 10 negative emotions on five points scale. Collected excerpts with French audio tracks are available online with their averaged assessed  scores\footnote{http://nemo.psp.ucl.ac.be/FilmStimuli}. 

Almost all research work published in the field of multimedia content analysis and emotions has used its own individually developed corpus. Table~\ref{tab:litReview} provides and overview of key examples of such work, specifying details of the affect categories used and the modalities of the video (i.e., audio and/or video) that were used to carry out the analysis. Table~\ref{tab:litReview} makes it possible to compare the corpora that were used to evaluate these techniques and the results. In the following, we provide additional details on this work, by discussing specific examples in greater depth.

Wang and Cheong~\cite{1637510} created and annotated a data set consisting of 36 full length Hollywood movies having 2040 scenes. Three annotators watched the movies and reported their emotions specifying Ekman basic emotion labels~\cite{citeulike:775844} for every scene. Only 14\% of the scenes received double labels and the rest only received single emotional labels from their three annotators.

Hanjalic and Xu~\cite{hanjalicxu:2005} used excerpts without annotations from the movies ``Saving Private Ryan'' and ``Jurassic Park 3'' and two soccer matches in their study. Irie et al.~\cite{5571819} used 206 selected emotional scenes out of 24 movies.  A total of 16 students annotated these scenes using eight Plutchik basic emotions: ``joy", ``acceptance", ``fear", ``surprise", ``sadness", ``disgust", ``anger", and ``anticipation"~\cite{citeulike:8791184}. The annotators first watched the videos and then reported how much they felt each of these emotions on seven-point scale. The emotional labels were assigned to the selected scenes only if more than 75\% of annotators agreed on them, otherwise the neutral label was assigned to the movie scene.
M. Xu et al. \cite{Xu:2008:HMA:1459359.1459457} used selected scenes from eight movies containing 6201 shots totaling 720 minutes. The videos were manually labeled by five emotions: ``fear", ``anger", ``happiness", ``sadness", and ``neutral", spanning the arousal dimension in three levels and valence in two levels.

%Mohammad: Could you check out what happened here. We lost something it seems!
Soleymani et al.~\cite{citeulike:7927300} used 21 full length commercially produced movies. One annotator annotated the movies continuously using an annotation tool which was recording the coordinates of mouse on valence and arousal plane on every click. The annotator reported his emotion at every moment he felt a different emotion while watching the movies.

Teixeira et al.~\cite{citeulike:8638639} used selected excerpts from 24 movies. They first segmented the movies into short clips (M=112s), and showed them to 16 participants. Participants rated the movies using SAM Manikins \cite{citeulike:2239554} on a seven-point scale; 346 clips, 10h 26 min in total, were chosen to span arousal, valence, dominance space.

Demarty et al.~\cite{DBLP:conf/eccv/DemartyPGS12} created a benchmark consisting of 18 Hollywood movies for violence detection. Although the movies are not annotated directly with emotional terms, we include this example here since depiction of violence elicits a variety of strong emotions. The data set is annotated by seven annotators on shot level.

%There are two types of intrinsic emotional databases, acted and spontaneous. Despite the existence of large number of studies on emotional expressions and their databases, there are only few databases which includes spontaneous emotions. One of the notable databases with spontaneous reactions is the Belfast database (BE) created by Cowie et al. \cite{citeulike:7921301}. The BE database includes spontaneous reactions in TV talk shows. Although the database is very rich in body gestures and facial expressions, the variety in the background and quality makes the data a challenging dataset of automated emotion recognition. The posed databases are usually recorded with a constant background (blue or green curtain) with a fixed camera and lighting. There are also different acted dataset containing faces and gestures which are recorded for facial or gesture recognition studies \cite{gunes06, 5571821, citeulike:7921301}. 

\subsection{Open Issues with existing corpora}
\label{sec:openIssues}
We end this section with a summary of the limitations of currently existing corpora in the form of a catalogue of open issues. 

\textbf{Users' mood and context}: We have pointed out that the same viewer can experience different emotions in response to the same stimulus depending on the context. The importance of the influence of mood for viewer affective response to video is relatively uncontroversial. For example, it is not strange or surprising when someone remarks, ``I am not in the mood to watch that movie today."
Our survey has revealed that the assumptions and methodology adopted by existing work is inconsistent with the importance of context for affective reactions. Researchers often fail to emphasize controlling the context and conditions in which annotations are collected from users, or disregard the issue of context entirely when designing experiments and developing data sets. Ignoring or suppressing context introduces risk into affective video indexing research: a system that does not  take into account the high degree of variability that characterizes naturally occurring contexts in which video is consumed may not be able to respond appropriately to user needs in real-world situations. As we will discuss further in Section~\ref{sec:ourdatasets}, there are a wide variety of contextual dimensions with a significant effect on the emotions that viewers feel in response to a video, including time of the day, temperature, mood and social context.

%These factors are either absent or tried to be suppressed by controlling the experimental conditions in the existing corpora. 
\textbf{User variability}: Beyond contextual factors, most of the existing research on affective video characterization has assumed reactions to be homogeneous across viewers, e.g.~\cite{1637510}. In some cases, the assumption of a single, obvious affective reaction from viewers is so strong, that affective video analysis is carried out, without collecting any user annotations at all, e.g.,~\cite{hanjalicxu:2005}. In most cases, however, assuming that everyone will react in the same way when watching a particular video is a strongly limiting assumption, that contradicts our intuition that the subjective nature of affect includes a strongly individual dimension. Corpora that allow both the personal and general dimensions of affective reactions to be explored have greater potential in helping to advance algorithm development in a direction that will best cover the needs of the full spectrum of possible users.

\textbf{Representative sampling}: In order to model affective responses that vary over context and across videos, affective video corpora are needed that include a large number of responses collected from a very large and representative population. However, the number of viewers and their feedback are often limited by our experimental setting and resources. In order to carry out research within the practical constraints of the real world, methods for creating affective video indexing corpora must be both effective---resulting in useful, high-quality corpora---and also efficient with respect to both the time spent and the expense incurred in the development process.  

These open issues constitute three dimensions that inform our proposal of guidelines for corpus development for affective video indexing and will steer the development of new corpora to avoid the shortcomings of existing ones. In the next sections, we first introduce the proposed guidelines for affective video corpora and then we discuss how corpora that we have developed have moved progressively towards addressing these limitations.

\section{Guidelines for Affective Video Indexing Corpora}
\label{sec:specs}

In this section, we present a set of corpus development guidelines for affective video indexing. The guidelines are informed by the ground that we have covered thus far, i.e., understanding of emotions and affective response from psychology and techniques available to record it, and also by the general types of multimedia context analysis algorithms that we expect that researchers will be developing with the data sets. We also take into account the limitations of the currently existing corpora, just discussed. From this information, three dimensions emerge that are critical to take into consideration when developing corpora for affective video indexing.

{\bf Context of viewer emotional response}:
Emotional response is complex, and arises not just from the video, but from the context of the video. We consider context to be what the viewer was exposed to before and after the part of the video for which we are interested in the affective impact. Context also includes the people with whom the viewer is watching the video and the viewer's underlying mood and physical state. The complexity cannot be completely controlled, but its impact can be minimized by very explicitly planning the set up in which viewers are exposed to videos. An evaluation protocol should be included that describes exactly what the annotators were asked to do. The protocol ensures that the annotation situation is reproducible should it ever be necessary/desirable to extend the annotations. What is important is to remain firmly focused on how the task is defined so that it is clearly understood that we are trying to predict affective impact on the viewer. Modeling of affect expressed within the video is admitted. However, it should be understood that this is only used as a bridge to infer the ultimate impact on the viewer. It should be clearly stated which parts of the emotional response process, for example, the affective and cognitive components vs. the conative and physiological components. The implications of ignoring the other components should be taken into account. 

The formulation of the way in which self-reported emotions are elicited should control for the impact of video before and after the target segment. This includes showing enough of the video. It is important to realize that entertainment video ``works" exactly because it takes us as viewers through alternations of mood, or expresses more than one mood at once. Depending on how the video is split up for mood elicitations different (or impartial) viewer responses can be expected. Good handling of context also involves gathering information on the users' underlying mood and physical state. 

{\bf Personal variation among viewers}:
Personal variation among viewers has a variety of sources. Some of the personal variation can be dealt with by careful handling of context, as mention above. Classical demographic differences are another source of variation. It is important that the target group be defined clearly (e.g., children) so that any existing limitations on user-to-user variability can be as well understood as possible. Narrowing the target group to a very small demographic (e.g., university students in their twenties) should be understood to limit the general applicability of the annotations gathered. 

Personal reactions vary according to personal topic preference. It is important to abstract away from topic or the topical interest of viewers: this can be accomplished by using a well balanced data set. Alternately, during data set design, a decision can be made to focus on one particular topic or style of data, which is significant enough to merit study. In any case, the corpus should be as multifunctional as possible: for example, involve enough users so that not only can universal reactions be studied, but also it is possible to study the reactions of different clusters of users that have similar responses.

{\bf Effectiveness and efficiency}:
%(planned handling of the corpus development process):
In practice, evaluation corpora are always developed under limitations of resources including person power and time. It is important to carefully plan how the corpus development process is handled. Decisions how to most effectively allocate limited resources have a critical effect on the usefulness of the corpus. As much as possible, such decisions should not be made in an arbitrary manner or during the actual process of gathering annotations for the corpus. In order to avoid unnecessarily jeopardizing the usefulness of the corpus, design decisions should be informed by the overall scenario or scenarios for which the data set is being developed. An overall scenario will assure that the type of affective response collected is appropriate for domain. For example, in the case of sports video the expected affective response pattern will be different than that for television talk shows. Further, the fact that resources are limited means that there is a trade-off between the resolution of the annotations that can be collected and the amount of content that participants can annotate. More sophisticated systems, e.g., PANAS, will lead to very high quality annotations. However, less people are eager to participate in the studies in which the process of response formulation is tedious or otherwise burdensome. Understanding the underlying use scenario, will make it possible to make informed decisions about trade-offs during the corpus design process.

An overall scenario is also helpful, should it become necessary to make further design decisions during the course of corpus development, e.g., decisions how to most effectively use limited resources in the case of unexpected loss of time or budget. Also, if researchers want to reuse the data set later, they have an idea of which uses are appropriate and which uses overstretch what the data set is designed to do. It is also useful to take into account the kind of multimedia content analysis that will be developed and the evaluation measure that will be used. However, it is of critical importance that the corpus be designed to reflect human affective reactions and not be biased to the specific algorithms, or types of algorithms, whose development it is intended to support.

\vspace{-10pt}
\section{Example Corpora}
\label{sec:ourdatasets}
% Statement of the tasks, Why and then what and finally so what
To demonstrate the application of these guidelines we now turn to discuss concrete examples. Three affective video corpora have been developed using three approaches of increasing sophistication.
%, which progressively approach the ideal benchmarking corpus.
%These datasets are presented in the order they were developed in time. 
The lessons learned from each corpus development experience were used to improve the next corpus. 
As such, the corpora represent a progression that moves towards an ideal corpus for affective video indexing.
The annotations for the first dataset were gathered in a laboratory setting. The second dataset was annotated with user affective responses gathered via a Web-based online platform, and the third dataset includes affective responses gathered using an online crowdsourcing platform. 
%In the development of every dataset we tried to address the problems we faced in the previous experiment. 
The general characteristics of the corpora are presented in Table~\ref{tab:corp_over} for easy reference. The next three sections discuss each in turn and the section includes with additional comparative discussion.

\begin{table}[ht]
\caption{The specifications of the three developed corpora.}
\vspace{-10pt}
\label{tab:corp_over}
\begin{center}
\begin{tabular}{{p{2.4cm}p{5.6cm}}}
\hline
\noalign{\smallskip}
\hline
\noalign{\smallskip}
\multicolumn{2}{c}{\textbf{Laboratory-based responses to movie scenes}}\\
\noalign{\smallskip}
\hline
\noalign{\smallskip}
\textbf{Nr. of participants} & 10 participants, 3 female, 7 male\\
\noalign{\smallskip}
\textbf{Self-reports} & free choice words, continuous arousal and valence using sliders using SAM Manikins\\
\noalign{\smallskip}
\textbf{Stimuli} & Short movie scenes from Hollywood movies\\
\noalign{\smallskip}
\textbf{No of videos} & 64\\
\noalign{\smallskip}
\textbf{Selection method} & manual\\
\hline\hline
\noalign{\smallskip}
\multicolumn{2}{c}{\textbf{Web-based responses to movie scenes}}\\
\noalign{\smallskip}
\hline
\noalign{\smallskip}
\textbf{Nr. of participants} & 42 participants, 15 female, 27 male\\
\noalign{\smallskip}
\textbf{Self-reports} & forced choice from 7 words, arousal and valence on 9 points scale using SAM Manikins\\
\noalign{\smallskip}
\textbf{Stimuli} & Short movie scenes from Hollywood movies\\
\noalign{\smallskip}
\textbf{No of videos} & 155\\
\noalign{\smallskip}
\textbf{Selection method} & manual\\
\hline\hline
\noalign{\smallskip}
\multicolumn{2}{c}{\textbf{Crowd-based responses to travelogue videos\footnote{This data set is available to researchers as a part of MediaEval benchmarking datasets. Please, contact MediaEval contacts that can be found at http://www.multimediaeval.org/}}}\\
\noalign{\smallskip}
\hline
\noalign{\smallskip}
\textbf{Nr. of participants} & 32 participants, 11 female, 18 male\\
\noalign{\smallskip}
\textbf{Self-reports} & forced choice from 11 words, boredom and like-dislike rating on 9 points scale\\
\noalign{\smallskip}
\textbf{Stimuli} & travelogue videos from ``The Big Travel Project''\footnote{http://www.mynameisbill.com/}\\
\noalign{\smallskip}
\textbf{No of videos} & 125\\
\noalign{\smallskip}
\textbf{Selection method} & The whole series\\\hline
\end{tabular}
\end{center}

\end{table}

\subsection{Movie scenes annotated in a laboratory}
\label{subsec:labDataset}
\subsubsection{Emotional videos}
The first corpus that was developed is comprised of emotional movie scenes suitable for emotion elicitation and characterization. The affective annotations were gathered via an experiment in which short video clips were shown to participants in a laboratory setting and their physiological responses and emotional self-reports were recorded. Self-reporting included both reporting words, that participants were free to choose themselves, and also reporting continuous arousal and valence using sliders using SAM Manikins. 

The physiological responses recorded were peripheral physiological signals generally used for assessing emotions, specifically: Galvanic Skin Response (GSR), Blood Volume Pulse (BVP), which provides heart rate, respiration pattern, and skin temperature. In order to capture facial muscle activity, we also recorded electromyograms (EMG) from the Zygomaticus major and Frontalis muscles. The physiological responses from 8 participants out of 10 were analyzed; signals from two participants were discarded due to technical problems. Subsequently, we calculated audio and visual content features from the videos and studied their correlation with both emotional responses and physiological changes. The GSR, BVP, EMG from Frontalis and Zygomaticus muscles were found to have significant correlation with arousal whereas only facial EMG was among the highly correlated features with valence, i.e., facial expressions were better indicators for valence compared to peripheral physiological signals. EMG from Zygomaticus muscles was found to be correlated with key lighting in the movie scenes and features extracted from BVP were found to be correlated with shot length variation. More detail on the results and analysis of the physiological responses can be found in~\cite{Soleymani:2009:IJSC}.

Due to the limited time that a participant can spend in each session, a relatively small set of videos, 64 clips from eight movies, were chosen and shown in two sessions. To create this video dataset, we selected video scenes from movies chosen either by using movies selected by similar studies (e.g., \cite{1637510,citeulike:8017077,hanjalicxu:2005}), or by choosing recent popular movies. The set of movies included four major genres, namely, drama: The Pianist and Hotel Rwanda; horror: The Ring (Japanese version) and Days Later; action:  Kill Bill Vol. I and Saving Private Ryan; comedy: Mr. Bean's Holiday and Love Actually. These four main genres were selected based on a previous study by Wang and Cheong~\cite{1637510}. Although the main genres of the movies were limited to these four, no constraints were imposed on secondary genres, meaning that the movie set also included aspects of, e.g., romance, thriller, sci-fi, history.
%Video clips used for this study are extracted from the list given in Table \ref{tab:8films}. 
The scenes that were selected, eight for each movie, had durations of approximately one to two minutes each and contained an emotional event (as judged by the first author). The complete list of the scenes with editing instructions and descriptions is available online\footnote{http://cvml.unige.ch/movieList}.
%  
%\begin{table}
%\renewcommand{\arraystretch}{1.3}
%\caption{Movies contained in the first corpus, organized by genre. Index number assigned to the movie is shown in parentheses after each title.}
%\begin{center}
%\begin{tabular}{p{4cm}|p{4cm}}
%\hline\hline
%\noalign{\smallskip}
%\textbf{Drama movies} & \textbf{Comedy movies}\\\hline
%The Pianist (6), Hotel Rwanda (2) & Mr. Bean's Holiday (5), Love Actually (4)\\\hline
%\noalign{\smallskip}
%\textbf{Horror movies} & \textbf{Action movies}\\\hline
%The Ring (Japanese version) (7), 28 Days Later (1) & Kill Bill Vol. I (3), Saving Private Ryan (8)\\\hline
%\end{tabular}
%\end{center}
%\label{tab:8films}
%\end{table}

\begin{figure}[h!]
\begin{center}
\includegraphics[width=0.5\linewidth]{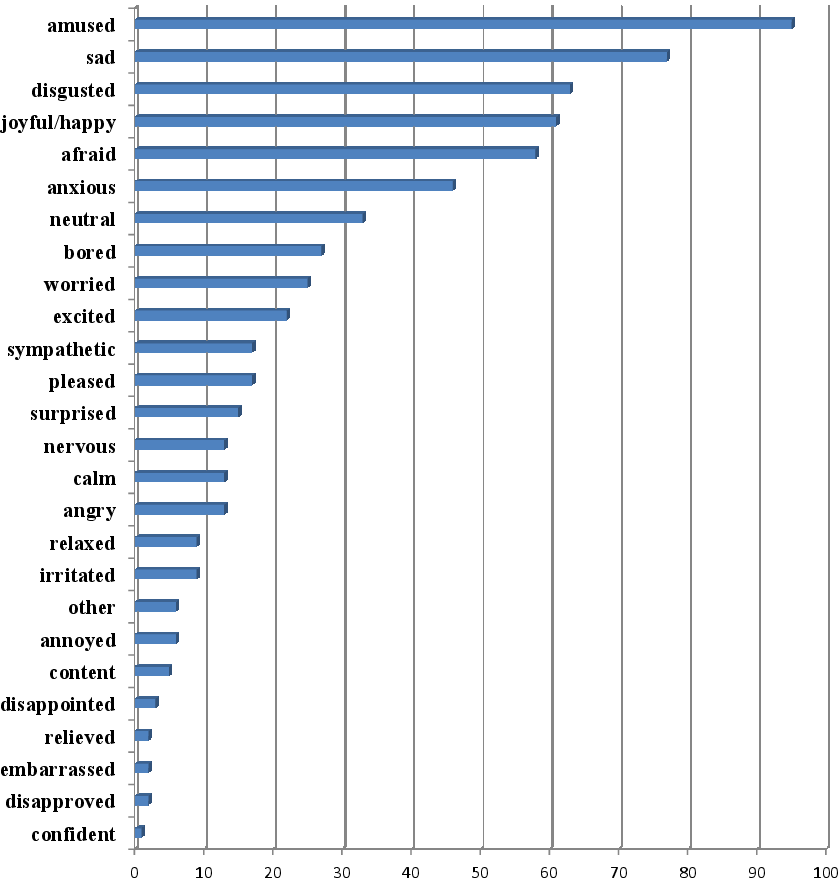}

\caption{Total number of keywords reported by 10 participants in response to the 64 video clips of movie scenes in the first corpus.}
\label{fig:labels}
\end{center}

\end{figure}

\subsubsection{Analysis of assessments}
The annotations were collected from ten (three female and seven male) participants ranging in age from 20 to 40 years ($M=29.3,SD=5.4$). 
The difference between arousal and valence scores given by the participants to all the videos was studied by means of a multi-way ANalysis Of VAriance (ANOVA), which was performed on arousal and valence scores considering three factors: the video scenes, the participants, and the order in which the videos were shown to the participants during sessions.
The effect of the order in which the videos were presented to the users on the user response was not significant. However, there was a significant difference on average valence scores between different participants ($F(9)=18.53, p<1\times10^{-5}$) and different videos ($F(63)=12.17, p<1\times10^{-5}$). There was also a significant difference on average arousal scores between different participants ($F(9)=19.44, p<1\times10^{-5}$) and different videos ($F(63)=3.23, p<1\times10^{-5}$). 
%Note4MS:  I think we need to change this sentence to read:
%These differences can be considered to originate not only from affective response to the content of the movies, but also from different personal experiences and memories concerning different movies, as well as participants' mood and background.
These differences can be attributed to different personal experiences and memories concerning different movies, as well as participants' mood and background.

%\begin{figure}[ht]
%\begin{center}
%\includegraphics[width=8cm]{figs/GenevaDistrARVAL.eps}
%\end{center}
%\caption{The distribution of different movie scenes on arousal and valence plane. Average arousal and valence are shown. Different numbers represent different movies (see Table \ref{tab:8films}).}
%\label{fig:VAdist}
%\end{figure} 

%The distribution of average arousal and valence scores are shown in Fig.~\ref{fig:VAdist}. The numbers that represent the movie scenes are the index codes associated with the movies in Table \ref{tab:8films}. The variance along the valence dimension was observed to increase with arousal. This observation is consistent with the findings of \cite{citeulike:8878318} in which arousal and valence scores in response to International Affective Picture System (IAPS) and International Affective Digital Sounds (IADS) showed a parabolic or heart shape distribution.

The development of this corpus provided an important lesson about the personal nature of user-reported affective response and the importance of carefully designing the method for collecting self-reported affective keywords from experiment participants. During the experiments the participants remarked that it was difficult for them to come up with emotion words when watching a video scene. In the end, there was a very low level of consensus among the words that they chose.  %The overall set of keywords chosen by participants was  biased towards frequently used basic emotions, e.g., anger, was not very common compared to non-basic emotions, e.g., amusement (see Fig. \ref{fig:labels}).
The overall set of keywords chosen by the participants did not include a high number of instances of basic emotions, e.g., anger, was not very common compared to non-basic emotions, e.g., amusement (see Fig. \ref{fig:labels}). These observations led to the lesson that giving users complete freedom of choice of response is not particularly helpful in isolating those common aspects of affective response. Instead, it is easier for participants, and yields more stable results, if participants choose from a list of choices. However, the list should not be blindly adopted from the literature, but should be carefully developed for a particular setting using exploratory experiments with participants. 
%Overall, the more participants that contribute affective response reports, the higher the consensus will be.

%The free choice keywords provided reported by the participants showed us that the free choice self reports are not appropriate for classification and categorization purposes (see Fig. \ref{fig:labels}). The usage of some of the frequently used modal emotions, e.g. anger, was not very common compared to non-modal emotions, e.g., amusement. This led not to use the free choice self reports and preselect a subset of keywords in the following experiments. 

\subsection{Web-based annotated movie scenes dataset}
\label{subsec:webBased}

\subsubsection{Emotional videos}
The development of the second corpus targeted the involvement a larger set of participants. First, a user study was conducted to narrow down the selection of videos to be used as stimuli. This time a more efficient forced-choice self-reporting was used. In order to find videos eliciting emotions from the whole spectrum of possible emotions, a user study was conducted to annotate a set of manually preselected movie scenes. The dataset is drawn from 16 full length Hollywood movies (see~\cite{5349526} for the full list). % which are listed in Table \ref{tab:films} (mostly popular movies). 
To create this video dataset, we extracted video scenes from movies selected either according to similar studies (e.g.,~\cite{1637510,citeulike:8017077,hanjalicxu:2005,Soleymani:2009:IJSC}), or from recent famous movies. A set of 155 short clips, each about one to two minutes long, were manually selected from these movies to form the dataset.

%Table gives the titles of the movies and online videos from which the emotional video clips were chosen.
% \begin{table}
% \renewcommand{\arraystretch}{1.3}
% \caption{The video clips were extracted from the listed movies}
% \begin{center}
% \begin{tabular}{p{4cm}|p{4cm}}
% \hline\hline
% \noalign{\smallskip}
% \textbf{Drama movies} & \textbf{Comedy movies}\\\hline
% The pianist, Hotel Rwanda, Apocalypse now, American history X, Hannibal & Man on the moon, Mr. Bean's holiday, Love actually\\\hline
% \noalign{\smallskip}
% \textbf{Horror movies} & \textbf{Action movies}\\\hline
% Silent hill, 28 days later, The shining & Kill Bill Vol. I, Kill Bill Vol. II, Platoon, The thin red line, Gangs of New York\\\hline
% \end{tabular}
% \end{center}
% \label{tab:films}
% \end{table}

%\begin{figure}
%\begin{center}
%\includegraphics[width=8cm]{figs/snapshot.eps}
%\caption{A snapshot of the affective annotation platform.}
%\label{fig:snapshot}
%\end{center}
%\end{figure} 

%A web-based annotation system has been launched to assess participants' felt emotion. 
A Web-based annotation system was developed and deployed in order to collected participants' self-reported affective responses.
To use this system, experiment participants sign up and provide personal information including gender, age, and email address. The system also collects information such as cultural background and origin that is used to build a profile of the experiment participants. Providing this information is optional. 
%Fig.~\ref{fig:snapshot} shows a screenshot of the assessment interface where a video clip is being shown. 
After watching each video clip, participants express the emotion that they felt using arousal and valence, quantized in nine levels. Participants also choose the emotional label best reflecting the emotions that they feel upon watching the clips. The emotion labels are ``afraid", ``amused", ``anxious", ``disgusted", ``joyful", ``neutral", and ``sad". These labels were chosen based on an assessment of the most frequent labels that were used by participants to describe their emotional responses during the development of the first data set (see Section~\ref{subsec:labDataset}).
Recall that for the first data sets, we asked participants to freely express their emotions, as elicited by movie scenes, with words. The emotional keywords used for the second dataset described in this section are the ones which were used most frequently by participants contributing to the first dataset~\cite{Soleymani:2009:IJSC} (see Fig. \ref{fig:labels} for the full list of words contributed during the collection of the first dataset). 
%Note4MS: I don't quite understand what we are saying here. You are referring back to the previous experiment, but the link to this experiment escapes me.
%During the initial experiments, the laboratory based experiment, we asked 10 participants to freely express their emotions, elicited by movie scenes with words. These emotional keywords were the ones which appeared more frequently~\cite{Soleymani:2009:IJSC} (see Fig. \ref{fig:labels}). Note that they roughly correspond to the six basic or modal ``Ekman's emotions''~\cite{citeulike:775844}.

\subsubsection{Analysis of the self-reports}

\begin{figure}[ht]
\begin{center}
\includegraphics[width=0.5\linewidth]{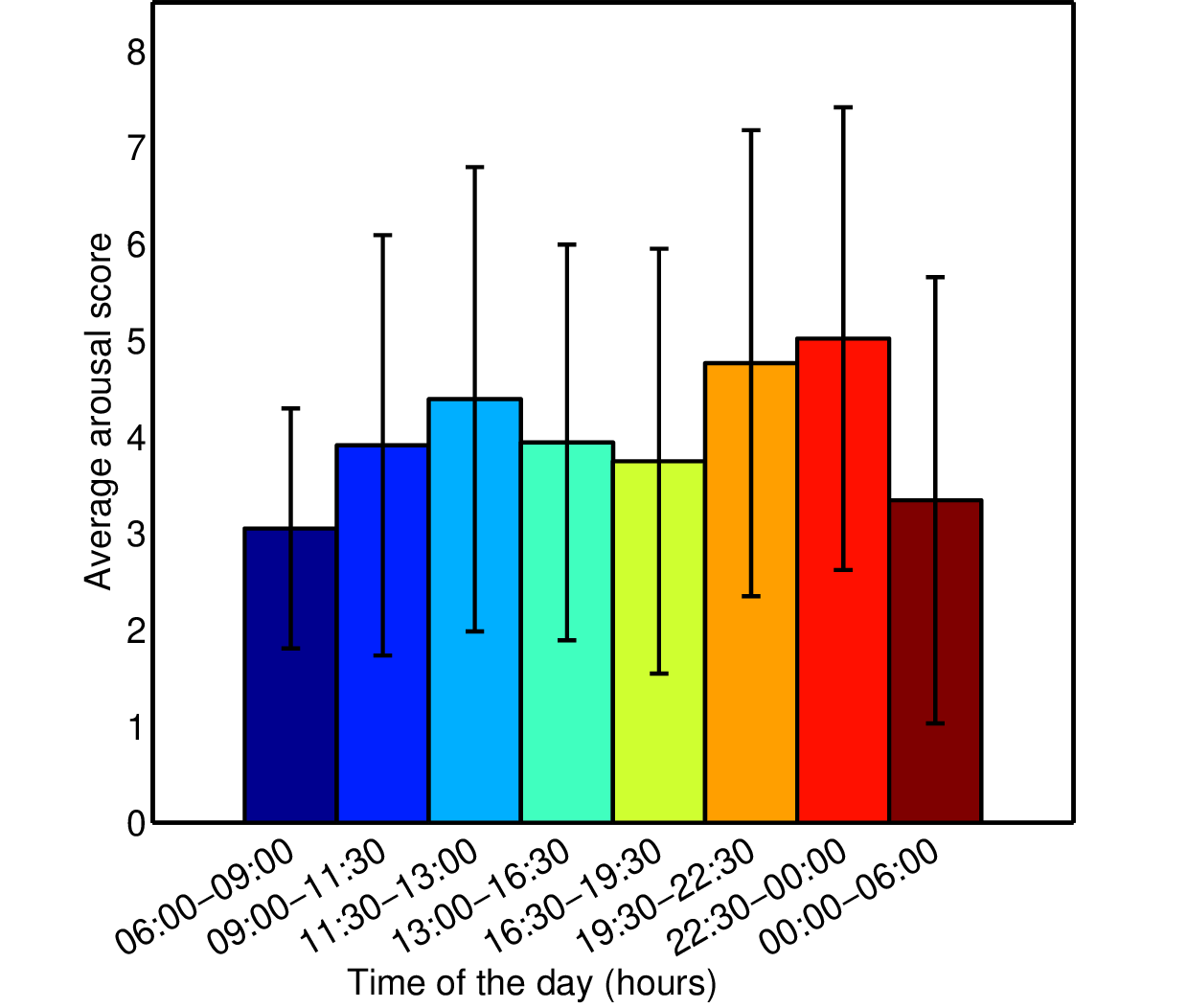}

\caption{Average arousal scores in different times of the day. Error bars represent the standard deviation of the ratings.}
\label{fig:arousalTime}
\end{center}

\end{figure}

Initially, 82 participants signed up to annotate the videos. From these 82 participants, 42 participants annotated at least 10 clips. Participants were from 20 to 50 years old ($M= 26.9,~SD = 6.1$). Out of the 42 participants, 27 were male and 15 were female with different cultural backgrounds living in four different continents.  
%The results of a multi-way ANOVA on arousal scores as the dependent variable and participant, video clip, and time of day as effects showed that the average arousal scores have a significant difference for different participants ($F(41)=3.23, p<1\times10^{-5}$), video clips ($F(154)=5.35, p<1\times10^{-5}$) and time of day ($F(7)=2.69, p<0.01$). 
We used a linear mixed model to test the effect of the time of day on the arousal and valence scores given to the clips. Participant and video clip were considered as random effects and  the time of day as fixed effect. The results showed that the average arousal scores are significantly different in different times of the day, i.e., the ANOVA test showed that the coefficients corresponding to different times of the day in the mixed linear model were significantly different from zero   ($F(7)=2.8, p<0.007$). We did not find the time of day to have significant effect on valence scores.
A day was divided into eight time interval, early morning (6:00 to 9:00), morning (9:00 to 11:30), noon (11:30 to 13:00), afternoon (13:00 to 16:30), evening (16:30 to 19:30), late evening (19:30 to 22:30), night (22:30 to 24:00) and after midnight (00:00 to 6:00). The average arousal scores in different time periods are shown in Fig. \ref{fig:arousalTime}. The average arousal scores given to all videos increases from early in the morning until noon. Then it decreases until it bounces back for late evening and night. The effect of circadian rhythm on self-reported arousal levels reflects the impact of context.  
Female participants on average gave higher arousal scores to the videos. %(see Fig. \ref{fig:arousalGender}). 
%Note4MS: I think that somewhere in this paragraph you want to add that these were the same sort of gender differences as were observed by Rottenburg (and that this confirms that the Web-based experiments is picking up on the same trends as lab-based experiments)
A Wilcoxon rank sum test showed that the difference between female and male participants' arousal scores was significant  ($p = 3\times10^{-16}$). These results are in line with the previous findings, e.g.,~\cite{citeulike:8017077}, which showed women report stronger emotions than men in response to the same stimuli.

%Note4MS: I think we decided to eliminate this paragraph
%Studying the variance of valence is more tricky since the dataset has roughly a balanced set of pleasant and unpleasant videos. Looking at the average valence scores, no significant difference can be observed between the ratings given by different gender groups or in different time intervals. 

Using the Web-based annotation system had a clear advantage because it increased the number of users from which we were able to collect annotations. 
%Note4MS: Could you check to make sure this sentence is right?
We were able to collect enough annotations so that it was meaningful to analyze the variance of the reported arousal scores over times of the day. Also, the Web interface meant that the participants could annotate more video than what was possible in two lab sessions of limited length. Additionally,  the development of this corpus led to an important lesson about the context of user annotations. Using the Web interface meant that the environment of the affective response was less controlled. The time of day had a significant impact on the response and it was important to record information about the influence of this factor.

% \begin{figure}[ht]
% \begin{center}
% \includegraphics[width=0.5\linewidth]{figs/gender_arousal_bars.png}
% \caption{Average arousal scores given by male and female participants.}
% \label{fig:arousalGender}
% \end{center}
% \end{figure}
% 

\subsection{Boredom prediction dataset}
\subsubsection{Crowdsourcing for affective annotation}
In order to reach a broader, more diverse, and larger population, a crowdsourcing platform, Amazon Mechanical Turk (MTurk)\footnote{http://www.mturk.com}, was used to gather annotations in the development of the third dataset. The third dataset, initially described in~\cite{Soleymani:CSE2010}, was developed with the aim of supporting research on video processing algorithms capable of predicting viewer boredom.  A video dataset has been gathered in the context of the MediaEval\footnote{http://www.multimediaeval.org} 2010 Affect Task for boredom prediction of Internet videos. Using MTurk we rapidly gathered self-reported boredom scores from a large user group that is demographically diverse and also represented our target population (Internet video viewers). Again, the forced choice emotional self-reporting methods were employed. Crowdsourcing practices leverage past work on recruiting subjects and conducting psychological experiments over the Web, which started in the nineties. A set of guidelines were identified by Reips~\cite{web_studies} to gather high quality responses from Internet users. Kittur et al. \cite{citeulike:2686454} showed that when appropriate measures were taken, crowdworkers rating of the quality of Wikipedia articles was comparable with that of experts.

For this work, we adopted a relatively simple, straightforward definition of viewer-experienced boredom. Boredom was taken to be related to the viewer's sense of maintaining focus of attention and is related to the apparent passage of time~\cite{citeulike:8847316}. Boredom is understood to be a negative feeling associated with viewer perceptions of the viewer-perceived quality (i.e., viewer appeal) of the video being low. 

The dataset selected for the corpus is Bill's Travel Project, a travelogue series called ``My Name is Bill'' created by the film maker Bill Bowles\footnote{http://www.mynameisbill.com}. The series consists of 126 videos between two to five minutes in length. This data set was chosen since it represents the sort of multimedia content that has risen to prominence on the Web. Bill's travelogue follows the format of a daily episode related to his activities and as such is comparable to ``video journals'' that are created by many video bloggers. %The results of analysis on video series such as ``Bill's Travel Project'' can extend to other video bloggers, and also perhaps to other sorts of semi-professional user generated video content. Because the main goal of this study was to study the effect of content related features, by using one series, the effect of high variance between content generated by multiple film makers in different genres was avoided. 

\subsubsection{Design of crowdsourcing task}
The third corpus that was developed once again increased the number of annotators and also introduced an even more sophisticated mechanism for context control. The affective responses for this corpus where collected using a large commercial crowdsourcing platform, Amazon Mechanical Turk (http://www.mturk.com). A crowdsourcing platform is an online labor market in which microtasks are offered by requesters and carried out by a pool of human users referred to as ``workers". Work in the area of human judgment and decision-making has revealed that there is no difference in the magnitude of the observed effects when experiments are performed using Mechanical Turk and when they are performed with a conventional pool of subjects~\cite{paolacci2010}. 
%{\color{dark} Other researchers have also discussed and used crowdsroucing for affective computing and affective video annotation and verified its relevance and significance \cite{riek2011guess, morris2011crowdsourcing,morency2011towards}.}

The crowdsourcing strategy used for the third corpus was designed based on the existing crowdsourcing literature, for example \cite{citeulike:2686454}, online articles and blog posts about crowdsourcing such as ``Behind the enemy lines'' blog\footnote{http://behind-the-enemy-lines.blogspot.com}, and also taking into account our past experience regarding collecting annotations in the Web-based experiment. A two-step approach was taken for our data collection. The first step was the pilot that consisted of a single micro-task or Human Intelligent Task (HIT) involving one video. This first HIT was used for the purpose of recruiting and screening MTurk workers as experiment participants. The second step was the main task and involved a series of 125 micro-tasks, one for each of the remaining videos in the collection. Workers were paid 30 US dollar cents for each HIT that they successfully completed.

The pilot HIT contained three components corresponding to responses that were required from the experiment participants that we recruited. The first section contained questions about the personal background (i.e., age, gender, cultural background). The second section contained questions about viewing habits: workers were asked whether they were regular viewers of Internet videos. The third section confirmed their seriousness by asking them to watch the video, select a word that reflected their mood at the moment, and also write a summary. The summary constituted a ``verifiable'' question, recommended by~\cite{citeulike:2686454}. The summary offered several possibilities for verification. Its length and whether it contained well-formulated sentences gave us an indication of the level of care that the worker devoted to the HIT. Also, the descriptive content indicated whether the worker had watched the entire video, or merely the beginning. A final question inquired if they were interested in performing further HITs of the same sort. In order not to directly reveal the main goal of the study to workers, the text box for the video summary was placed prominently in the HIT. 
%\subsubsection{Main Task}

The workers were chosen and qualifications were granted for the main task from the participants of the pilot by considering the quality of their description and answers. In the choice of workers, we also strove to maintain a diverse group of respondents. %The qualification was only granted to the participants who answered all the questions completely. The workers were invited to do the main study by sending them an invitation e-mail via their ID number on the MTurk platform. %The e-mail informed the users that our qualification was granted to them. Use of a qualification served to limit those workers that carry out the HIT to the invited workers. 
Each HIT in the main study consisted of three parts. In the first part, the workers were asked to specify the time of day. Also the workers were asked to choose a mood word from a drop down list that best expressed their reaction to an imaginary word (i.e., a nonsense word), such as those used in~\cite{Quirin2009500}. The mood words were ``pleased", ``helpless", ``energetic", ``nervous", ``passive", ``relaxed", and ``aggressive". The answers to these questions gave us an estimate of their underlying mood. In the second part, they were asked to watch the video and give some simple responses to the following questions. They were asked to choose the word that best represented the emotion they felt while watching a video from a second list of emotion words in the drop down list. The emotion list contained the Ekman six basic emotions~\cite{citeulike:775844} (namely, ``sadness", ``joy", ``anger", ``fear", ``surprise", and ``disgust") in addition to ``boredom", ``anxiety", ``neutral" and ``amusement". For this data set, we had little advance information concerning the emotions that we could expect the video content to trigger in viewers. For this reason, we chose the emotional categories to cover the entire affective space, as defined by the conventional dimensions of valence and arousal~\cite{citeulike:2642710}, supplemented by general information on emotions elicited by film~\cite{citeulike:8017077}. The emotion and mood word lists contained different items in order to avoid as much as possible that the experiment participants would strongly associate the two. Next, participants were asked to provide a rating specifying how boring they found the video and how much they liked the video, both on a nine point scale. % Then, they were asked to estimate how long the video lasted. Here, we had to rely on their full cooperation in order not to cheat and look at the video timeline. 
Finally, they were asked to describe the contents of the video in one sentence. %We emphasized the description of the video rather than the mood word or the rating, in order to conceal the main purpose of the HIT. Quality control of the responses was carried out by checking the description of the video and also by ensuring that the time that they took to complete the HIT was reasonable. 

\subsubsection{Analysis of the ratings}
Our pilot HIT was initially published for 100 workers and finished in the course of a single weekend. We re-published the HIT for more workers when we realized we needed more people in order to have an adequate number of task participants. Only workers with the HIT acceptance rate of 95\% or higher were admitted to participate in the pilot HIT. In total, 169 workers completed our pilot HIT, 87.6\% of which reported that they watch videos on the Internet. We took this response as confirmation that our tasks participants were close to the target audience of our research. Out of 169 workers, 105 were male and 62 were female and two did not report their gender. Their age average was 30.48 with the standard deviation of 12.39. The workers in the pilot HITs identified themselves by different cultural backgrounds from North America. Having such a group of participants with a high diversity in their cultural background would have been difficult in a conventional setting, i.e., without using the crowdsourcing platform. Of the 169 pilot participants, 162 had interest in carrying out similar HITs. Out of the interested group, 79 workers were determined to be qualified and were assigned our task-specific qualification within MTurk. This means only 46.7\% of the workers who did the pilot HIT were able to answer all the questions and had the profile we required for the main task. 

In total, 32 workers participated and also annotated more than 60 of the 125 videos in the main task HIT series. This means only 18.9\% of the participants in the pilot and 39.0\% of the qualified participants committed to do the main task HIT series seriously. Of this group of 32 serious participants, 18 were male and 11 were female with ages ranging from 18 to 81 ($M=34.9,SD=14.7$). %To evaluate the quality of the annotations, the time spent for each HIT was compared to the video length. In 81.8\% of the completed HITs, the working duration for each HIT was longer than the video length. This means that in 18.2\% of the HITs we have doubts if the workers fully watched the videos. This shows the importance of having workers with the right qualifications and trustworthy pool of workers in annotation or evaluation hits. Rejecting those HITs reduced the number of workers who carried out more than 60 videos in the main series of HIT to 25 from which 17 are male and 8 are female ages ranging from 19 to 59 ($M= 33.9,SD=11.8$). 

The following questions were asked about each video to assess the level of boredom. First, how boring the video was on nine-point scale from the most to the least boring. Second, how much the user liked the video on the nine-point scale and third how long the video was. Boredom was shown to have on average a strong negative correlation, $\rho= - 0.86$ with liking scores. %The time perception did not show a significant correlation for all users and it varied from 0.4 down to -0.27. Although positive correlation was expected from boredom scores and the perception of time seven participants' boredom scores have negative correlation with the time perception. 
The correlation between the order of watching the videos for each participant and the boredom ratings was also examined. No positive linear correlation was found between the order and boredom score. This means that watching more videos did not increase the level of boredom and, in fact, for two of the participants it lowered their reported boredom level. Additionally, the correlation between the video length and boredom scores was investigated. No positive correlation was found between the boredom scores and videos' duration. We can conclude that longer videos are not necessarily perceived as more boring than the shorter videos.

%To measure the inter-annotator agreement, the Spearman correlation between participants' pairwise boredom scores was computed. The average significant correlation coefficient was very low $\rho = 0.05$. There were even cases where the correlation coefficients were negative, which shows complete disagreement between participants. The low inter-annotators agreement reflects the personal taste in boredom perception. However, The rank for average boredom scores were robust among the extreme cases and reproducible with a subset of users.   

In order to obtain the dominant mood from the mood words, first the responses of each participant were clustered into the three hours time intervals. In each three hours interval the most frequent chosen mood word was selected as the dominant mood. After calculating the dominant moods, we found that using the implicit mood assessment none of the participants had the ``relaxed'' as their dominant mood. 

\begin{figure}[ht]
\begin{center}
\includegraphics[width=0.5\linewidth]{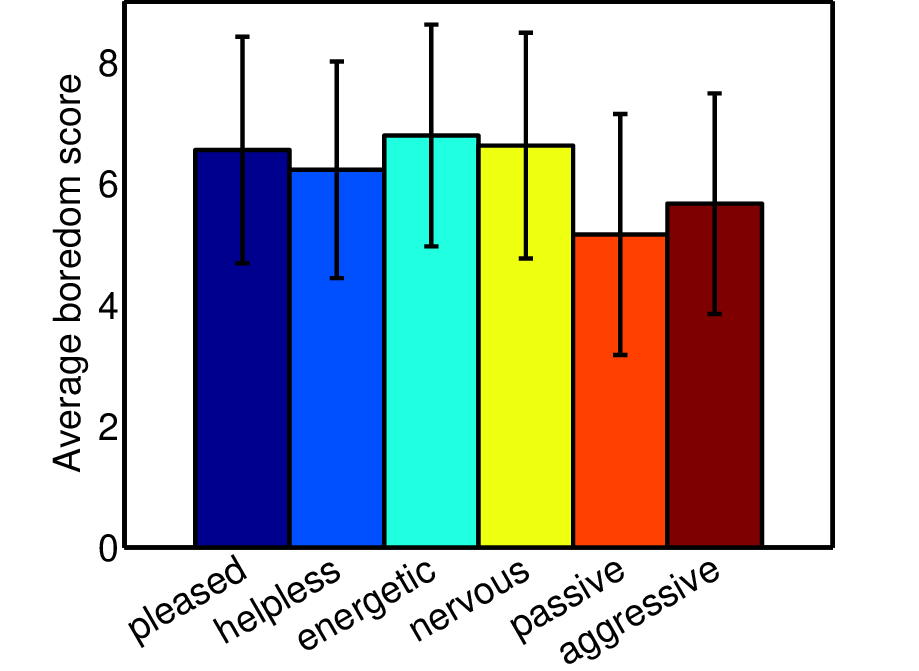}
\caption{Average dominant boredom scores reported by viewers experiencing different moods. Error bars represent the standard deviation of the ratings}
\label{fig:moodbars}
\end{center}

\end{figure}

The average boredom scores for different dominant moods are shown in Fig. \ref{fig:moodbars}. The boredom scores were, on average, lower, i.e., indicated that videos were \emph{more} boring, for viewers in a passive mood and higher, i.e., indicated that videos were \emph{less} boring, in an arguably more active mood such as ``energetic", ``nervous" and ``pleased". Moods were then categorized into  positive (``pleased", ``energetic", and ``relaxed") and negative, (``helpless", ``nervous", ``passive", and ``aggressive") categories. On average, participants gave higher ratings to videos when they were in positive moods. % (see Fig. \ref{fig:PNmoodbars}).
The statistical significance of the difference between ratings in positive and negative moods was examined by a Wilcoxon test and was found significant ($p=4\times10^{-8}$). 
The effect of the time of day and mood on boredom scores was investigated with a mixed linear model. The fixed effects were mood and time of day, the random effects were video and participant. Unlike the second data set, the effect of the time of day on boredom scores was not significant. This observation can be attributed to the difference between the nature of arousal and perceived boredom, arousal being more correlated with physiological state. Participants' mood had a significant effect on the ratings; the ANOVA test on the mood coefficients in the mixed linear model showed significant difference from zero $F(6)=5.75,p<3\times10^{-5}$). 
The analysis of annotations gathered in this dataset showed the importance of participants' mood which is often not assessed in affective and non-affective assessments.

From the development of this corpus we learned that it is possible to use commercial crowdsourcing to collect a large volume of user affective responses to video. The large number of participants made it possible to analyze sub-groups of participants with particular reactions. Affective response is personal, and varies from individual. However, looking at sub-groups of the population that pattern in the same way could make it possible to isolate the commonalities of viewer response. Variations in the context were addressed  by collecting information on the underlying moods of the participants. Even though crowdsourcing is relatively inexpensive, it is still important to plan resources carefully when designing a corpus that uses crowdsourcing to collect affective annotations. A trade-off needs to be made between more annotators, the number of videos annotated and the parts of the HIT design that verify engagement. Advanced planning was necessary to collect the annotations within a set amount of time, since workers may not immediately start working on a HIT once they have qualified. Also, it was necessary to have a large enough pool of workers available to work on the HIT, since some qualified workers do not return to work on the HIT after earning the qualification. In sum, the third corpus addressed all three dimensions of context, personal affective response, and tradeoffs of efficiency and effectiveness.

% \begin{figure}[ht]
% \begin{center}
% \includegraphics[width=0.5\linewidth]{figs/PNmoods_bar.png}
% \caption{Average boredom scores in positive and negative moods.}
% \label{fig:PNmoodbars}
% \end{center}
% \end{figure} 

\subsection{Discussion of the example datasets}
In this section, we provide additional discussion that compares the three data sets directly. As can be seen in Table~\ref{tab:corp_over}, data sets differ in the number of participants involved, the ways in which the participants were asked to self-report their emotional response, the nature of the stimuli (i.e., the video content), the number of videos in the data set, and the way in which the dataset was selected. 
Here, we focus our discussion on how the comparison of the data sets with regard to these dimensions sheds additional light on aspects of the guidelines for affective video corpora set out in Section~\ref{sec:specs}. The three data sets represent progressively more ambitious efforts to gather affective responses from a larger number of participants and for a larger number of videos. The final data set additionally moves from a hand selected set of videos to a set of videos more closely representative of video material that occurs ``in the wild". 

Comparison of the first two data sets provides important insight on how emotional self-reports can be collected for the purposes of corpus development for affective video indexing. In the first dataset, 10 participants contributed annotations for 64 videos. The size of this dataset was limited by the fact that a wide variety of physiological response signals were collected, in additional to self-reported affective response (cf. Section~\ref{subsec:labDataset}). The physiological response signals require setting up a large amount of measuring equipment for each response capturing session. The move to the Web-based data set allowed us to increase the number of participants from which data was collected. As can be seen in Table~\ref{tab:krip}, which reports inter-annotator agreement (Krippendorff's alpha) for the three data sets, moving from the conventional lab setting to the Web-based setting did not have a noticeable impact on the inter-annotator agreement. In sum, a comparison of the first two data sets confirms that it is possible to move beyond a conventional lab setting when collecting self-reported emotional response to videos and, instead, carry out collection using an on-line system, which does not require participants to be physically present in the lab.

%To measure the inter-annotator agreement, Krippendorff's alpha was calculated for the three developed dataset (see Table \ref{tab:krip}). The advantage of Krippendorff's alpha over other measures is that it can handle missing data and is applicable to bot nominal and ordinal categories. 

\begin{table}
\renewcommand{\arraystretch}{1.3}
\caption{Inter-annotator agreement (Krippendorff's alpha) for arousal, valence (ordinal scale)
%boredom and like-dislike ratings 
and emotion categories (nominal scale) on the three data sets: Lab-based (Lab), Web-based (Web) and Crowd-based (Crowd) are given.}
\begin{center}
\begin{tabular}{c|c|c|c}
\noalign{\smallskip}
\textbf{Database} & \textbf{Lab} & \textbf{Web} & \textbf{Crowd}\\\hline
\noalign{\smallskip}
\textbf{Valence} &0.49&0.47&-\\\hline
\textbf{Arousal} &0.13&0.22&-\\\hline
%\textbf{Boredom} &-&-&0.04\\\hline
%\textbf{Like-dislike} &-&-&0.05\\\hline
\textbf{Categories} & - &0.20&0.05\\\hline
\end{tabular}
\end{center}
\label{tab:krip}

\end{table}
%For the first experiment which was conducted in the laboratory environment, agreement was much higher for valence, $\alpha=0.59$, compared to arousal, $\alpha=0.13$. We did not calculate the agreement over the free choice, emotional keywords. For the Web-based experiment, similarly, the agreement was higher for valence compared to arousal. Looking at the agreement between these two experiments we can argue that online experiments yield similar results in terms of inter-annotation agreement which is a sign of reliability. This further verifies using online systems in lieu of laboratory based experiments.  The inter annotation agreement was lower for boredom and like-dislike scores for the third experiment. However, the nature of the content itself was also different. In the first two dataset, the stimuli consisted of pre-selected emotional scenes whereas in the database developed on crowdsourcing platform, the content was semi-professionally generated and was not pre-selected, i.e., we gathered annotations for the whole series of 125 videos.

It is important to note that we do not expect participants self-reported emotions to yield a high degree of inter-annotator agreement. As an example, consider the case of videos of sports games. Here, we can assume two groups of viewers, representing the supporters of the two opposing teams. Within a viewer-group, we expect viewers to report comparable responses to the same game events in the video content. However, we expect disagreement between the two groups on the emotional impact of a given event (i.e., a given viewer's response to a goal is determined by which team has scored the goal and which team that viewer is supporting). This underlying variation gives rise to weak overall agreement on emotional response. Moving from sports to other domains in which the trigger events in the video content are less well defined, the number of groups of viewers who pattern together with respect to their emotional response profiles can be expected to change. In particular, we would expect it to grow and for the response patterns to be less well defined. The field of affective video indexing is interested in investigating weak consistencies in viewer responses (i.e., that their distribution differs from random), or order to analyze their triggers across user groups to create better video retrieval and browsing systems.

Comparison of the second and third data sets provides important insights into the distribution of emotion response triggers in video. Recall that the second data set contained manually selected segments of movies that were chosen to create a even distribution in the response patterns that they evoked in viewers. The third data set, on the other hand, did not consist of pre-selected video, rather contains an entire series, i.e., a set of videos as they would occur ``in the wild". Both the second and the third data set contains emotion words that participants selected from a set list (i.e., forced-choice reporting of emotion words). Table \ref{tab:krip} shows that the inner-annotator agreement is higher for the manually selected video that for ``in the wild" video. This difference reflects the fact that ``in the wild" video might not be produced with the intention of evoking strong emotions or that the distribution of the emotion evoked by the video might not be evenly distributed across the categories (cf. the unevenness encountered by~\cite{riek2011guess} that is discussed in Section~\ref{sec:aviresponse}). In this case of the third data set, the video content is a collection of travelogues, which have a documentary as well as an adventure component to them. They can be expected to evoke emotions, but rather less frequently and also from a more limited set (e.g., ``happiness" and ``surprise" could be anticipated to be more frequent than ``fear"; an emotion such as ``anger" might never be triggered.) In sum, comparison of the second and third data sets confirm the importance of understanding the way in which the data collection used for corpus development matches the target data collection to which affective video indexing techniques developed using the data set will ultimately be applied.

We close this section with some final observations concerning emotion representations. Collecting emotional response in terms of free-choice emotional words gives the maximum freedom to participants to express their emotions. However, people are sometimes unable to come up with a good set of words and the analysis of such feedback is cumbersome for researchers. Forced-choice keyword based feedback is a good alternative in these situations. A narrow and well-designed set of choices can increase the consistency of participant responses. Note, however, that a fixed set of choices can only be used in situations in which it has already been established which emotions can be triggered.  Further, it is necessary to give careful consideration to how the emotional categories will ultimately be used in an application. It is not useful to collect a large number of fine resolution emotion labels from participants, if the application for which they are needed will only make use of several high level categories. Since a limiting factor in gathering emotional response feedback is the amount of time that participants are able to spend providing responses (cf. Section~\ref{sec:specs}), it may be more appropriate to ask participants to make a relatively ``easy" choice between a small number of categories, and collect annotations for a larger number of videos, rather than to gather very fine-grained labels. 

\section{Conclusion and Outlook}
%\section{Discussion and Recommendations}
\label{sec:conclusion}

This paper has addressed the development of corpora for research and evaluation in the area of affective video indexing algorithms and systems. 
%In particular, we have investigated the need in the field of affective video indexing for a rigorous benchmarking effort, evaluated the past experience in developing benchmarking resources and procedures and proposed new insights in this direction gained through the development of three new affective data sets, for which annotations were collected in a laboratory setting, on a web-based online platform and using an online crowdsourcing platform. 
We have proposed a set of guidelines that are intended to provide the research community with support in overcoming the deficiencies of the existing corpora for affective video indexing. Our investigation focused not only on the sources of the size- and scope-related limitations of the existing evaluation corpora, namely the difficulty of reliably collecting affective responses from test users and the need to reduce the variability in the noisy and subjective affective responses, but also on how other critical requirements related to the corpora development can be fulfilled. 

Our findings indicate that there are three key dimensions that need to be considered when developing corpora for affective video indexing research. The first dimension is context of viewer response. In particular, circadian rhythm and the mood of experiment participants have been shown to have significant effect on self-reported emotions. The second dimension is the personal differences. Significant differences have been observed between different participants' affective responses to the same content. The personal dimension emphasizes the importance of personalization and profiling strategies. Finally, efficiency and effectiveness are important factors to be taken into account. New methods for collecting annotations such as Web-based and crowdsourcing platforms offer improved opportunities to collect annotations in greater volumes and from wider diversity of users and a broader spectrum of contexts.
%Movie excerpts are mostly used for affective corpora development due to their effectiveness in inducing emotions in a large number of viewers.

These three dimensions form the basis for our proposed set of guidelines for affective indexing corpora. Three corpora are introduced which are developed with techniques that progressively approach an ideal corpus as defined by these guidelines. Forced-choice and simple emotional reporting methods have been employed in developing the corpora to decrease the effort of the participants and increase the efficiency of annotation collection and the ability of the annotations to reflect consistency and stability in self-reported affective responses. Larger populations of experimental participants can be reached with Web-based and crowdsourcing platforms, increasing both the diversity of the annotations collected and the ability of the corpus to reflect contextual and personal variation.

This paper has argued that high-quality corpora will help to push forward the state of the art in affective video indexing. In order to realize this predicted potential of affective video corpora, the next step is necessarily the development of additional corpora. If a multitude of corpora can be made available to the research community suitable to support research along the entire spectrum of possible affective video indexing applications, then researchers will have the necessary resources at their disposal to push affective video indexing into the next generation.

\ifCLASSOPTIONcompsoc
  \section*{Acknowledgments}
\else
  \section*{Acknowledgment}
\fi

The work of Larson and Hanjalic is supported in part by the European Community's Seventh Framework Program under grant agreement no. 287704 (CUbRIK). This work of Soleymani is supported by the European Research Area under the FP7 Marie Curie Intra-European Fellowship: Emotional continuous tagging using spontaneous behavior (EmoTag). 
%The research leading to these results has been performed in the frameworks of European Community's Seventh Framework Program (FP7/2007-2011) under grant agreement no. 216444 (PetaMedia). 
The authors would like to thank Bill Bowles for granting permission of using his video content for our boredom detection corpus.

\ifCLASSOPTIONcaptionsoff
  \newpage
\fi

\bibliographystyle{IEEEtran}
\bibliography{IEEEabrv,paper}

\begin{thebibliography}{10}
\providecommand{\url}[1]{#1}
\csname url@samestyle\endcsname
\providecommand{\newblock}{\relax}
\providecommand{\bibinfo}[2]{#2}
\providecommand{\BIBentrySTDinterwordspacing}{\spaceskip=0pt\relax}
\providecommand{\BIBentryALTinterwordstretchfactor}{4}
\providecommand{\BIBentryALTinterwordspacing}{\spaceskip=\fontdimen2\font plus
\BIBentryALTinterwordstretchfactor\fontdimen3\font minus
  \fontdimen4\font\relax}
\providecommand{\BIBforeignlanguage}[2]{{%
\expandafter\ifx\csname l@#1\endcsname\relax
\typeout{** WARNING: IEEEtran.bst: No hyphenation pattern has been}%
\typeout{** loaded for the language `#1'. Using the pattern for}%
\typeout{** the default language instead.}%
\else
\language=\csname l@#1\endcsname
\fi
#2}}
\providecommand{\BIBdecl}{\relax}
\BIBdecl

\bibitem{Kang:2003}
H.-B. Kang, ``Affective content detection using {HMMs},'' in \emph{ACM int'l
  conf. Multimedia}, 2003, pp. 259--262.

\bibitem{hanjalichighlights:2005}
A.~Hanjalic, ``{Adaptive Extraction of Highlights From a Sport Video Based on
  Excitement Modeling},'' \emph{IEEE Trans. Multimedia}, vol.~7, no.~6, pp.
  1114--1122, 2005.

\bibitem{Chan:2005}
C.~H. Chan and G.~J.~F. Jones, ``{Affect-based indexing and retrieval of
  films},'' in \emph{ACM int'l conf. Multimedia}, 2005, pp. 427--430.

\bibitem{SnoekFNTIR09}
C.~G.~M. Snoek and M.~Worring, ``Concept-based video retrieval,''
  \emph{Foundations and Trends in Information Retrieval}, vol.~4, no.~2, pp.
  215--322, 2009.

\bibitem{trecvid:2006}
A.~F. Smeaton, P.~Over, and W.~Kraaij, ``Evaluation campaigns and trecvid,'' in
  \emph{ACM Int'l Work. Multimedia Information Retrieval}, 2006, pp. 321--330.

\bibitem{trecvid.features}
A.~F. Smeaton, P.~Over, and W.~Kraaij, ``High-{L}evel {F}eature {D}etection
  from {V}ideo in {TRECV}id: a 5-{Y}ear {R}etrospective of {A}chievements,'' in
  \emph{Multimedia Content Analysis, Theory and Applications}, A.~Divakaran,
  Ed.\hskip 1em plus 0.5em minus 0.4em\relax Berlin: Springer Verlag, 2009, pp.
  151--174.

\bibitem{Arapakis:2008}
I.~Arapakis, J.~M. Jose, and P.~D. Gray, ``Affective feedback: an investigation
  into the role of emotions in the information seeking process,'' in \emph{ACM
  SIGIR}.\hskip 1em plus 0.5em minus 0.4em\relax ACM, 2008, pp. 395--402.

\bibitem{Lopatovska:2011}
I.~Lopatovska and I.~Arapakis, ``Theories, methods and current research on
  emotions in library and information science, information retrieval and
  human-computer interaction,'' \emph{Information Processing \& Management},
  vol.~47, no.~4, pp. 575 -- 592, 2011.

\bibitem{Picard95}
R.~W. Picard, ``Affective computing,'' MIT, Media Laboratory Perceptual
  Computing Section Technical Report 321, 1995.

\bibitem{hanjalic:2006}
A.~Hanjalic, ``Extracting moods from pictures and sounds: towards truly
  personalized tv,'' \emph{IEEE Signal Process. Mag.}, vol.~23, no.~2, pp. 90
  --100, 2006.

\bibitem{Lew:2006}
M.~S. Lew, N.~Sebe, C.~Djeraba, and R.~Jain, ``Content-based multimedia
  information retrieval: State of the art and challenges,'' \emph{ACM TOMCCAP},
  vol.~2, pp. 1--19, 2006.

\bibitem{Benini:2011}
S.~Benini, L.~Canini, and R.~Leonardi, ``A connotative space for supporting
  movie affective recommendation,'' \emph{IEEE Trans. Multimedia}, vol.~13,
  no.~6, pp. 1356--1370, 2011.

\bibitem{Thornley:2011}
C.~V. Thornley, A.~C. Johnson, A.~F. Smeaton, and H.~Lee, ``The scholarly
  impact of trecvid (2003--2009),'' \emph{J. Assn. Inf. Sci. Technol.},
  vol.~62, pp. 613--627, 2011.

\bibitem{rui:2000}
Y.~Rui, A.~Gupta, and A.~Acero, ``Automatically extracting highlights for {TV}
  baseball programs,'' in \emph{ACM int'l conf. Multimedia}, 2000, pp.
  105--115.

\bibitem{mitsu-dvd}
I.~Otsuka, S.~Shipman, and A.~Divakaran, ``\BIBforeignlanguage{English}{A video
  browsing enabled personal video recorder},'' in
  \emph{\BIBforeignlanguage{English}{Multimedia Content Analysis}}, ser.
  Signals and Communication Technology, A.~Divakaran, Ed.\hskip 1em plus 0.5em
  minus 0.4em\relax Springer US, 2009, pp. 1--12.

\bibitem{jones:2011}
G.~J.~F. Jones and C.~Hau~Chan, \emph{Affect-Based Indexing for Multimedia
  Data}.\hskip 1em plus 0.5em minus 0.4em\relax John Wiley \& Sons, Inc., 2012,
  pp. 321--345.

\bibitem{Janin:2010}
A.~Janin, L.~Gottlieb, and G.~Friedland, ``Joke-o-mat {HD}: browsing sitcoms
  with human derived transcripts,'' in \emph{ACM int'l conf. Multimedia}.\hskip
  1em plus 0.5em minus 0.4em\relax ACM, 2010, pp. 1591--1594.

\bibitem{4468714}
Z.~Zeng, M.~Pantic, G.~Roisman, and T.~Huang, ``A survey of affect recognition
  methods: Audio, visual, and spontaneous expressions,'' \emph{IEEE Trans.
  Pattern Anal. Mach. Intell.}, vol.~31, no.~1, pp. 39--58, 2009.

\bibitem{morency2011towards}
L.-P. Morency, R.~Mihalcea, and P.~Doshi, ``Towards multimodal sentiment
  analysis: harvesting opinions from the web,'' in \emph{ACM Int'l conf.
  multimodal interfaces}, 2011, pp. 169--176.

\bibitem{Biel2013}
J.~Biel and D.~Gatica-Perez, ``The youtube lens: Crowdsourced personality
  impressions and audiovisual analysis of vlogs,'' \emph{IEEE Trans.
  Multimedia}, vol.~15, no.~1, pp. 41--55, 2013.

\bibitem{morris2011crowdsourcing}
R.~Morris, ``The emergence of affective crowdsourcing,'' in \emph{ACM CHI '11
  Workshop on Crowdsourcing and Human Computation}, 2011.

\bibitem{citeulike:8222108}
W.~Wirth and H.~Schramm, ``{Media and Emotions},'' \emph{Communication research
  trends}, vol.~24, no.~3, pp. 3--39, 2005.

\bibitem{riek2011guess}
L.~D. Riek, M.~F. O'Connor, and P.~Robinson, ``Guess what? a game for affective
  annotation of video using crowd sourcing,'' pp. 277--285, 2011.

\bibitem{scherer2005}
K.~R. Scherer, ``{What are emotions? And how can they be measured?}''
  \emph{Social Science Information}, vol.~44, no.~4, pp. 695--729, 2005.

\bibitem{citeulike:2821039}
K.~R. Scherer, ``{Studying the emotion-antecedent appraisal process: An expert
  system approach},'' \emph{{Cognition \& Emotion}}, vol.~7, no. 3-4, pp.
  325--355, 1993.

\bibitem{ortony88emotion}
A.~Ortony, G.~L. Clore, and A.~Collins, \emph{{The Cognitive Structure of
  Emotions}}.\hskip 1em plus 0.5em minus 0.4em\relax {Cambridge University
  Press}, 1988.

\bibitem{citeulike:3014462}
D.~Sander, D.~Grandjean, and K.~R. Scherer, ``{A systems approach to appraisal
  mechanisms in emotion},'' \emph{Neural Networks}, vol.~18, no.~4, pp.
  317--352, 2005.

\bibitem{citeulike:8227534}
D.~Zillmann, \emph{{The psychology of suspense in dramatic exposition}}.\hskip
  1em plus 0.5em minus 0.4em\relax Lawrence Erlbaum Associates, Inc, 1996, pp.
  199--231.

\bibitem{citeulike:8227551}
A.~I. Nathanson, \emph{{Rethinking Empathy}}.\hskip 1em plus 0.5em minus
  0.4em\relax Lawrence Erlbaum Associates, Inc, 2003, ch.~5, pp. 107--130.

\bibitem{citeulike:8227621}
D.~Zillmann, \emph{{Empathy: Affect from bearing witness to the emotions of
  others}}.\hskip 1em plus 0.5em minus 0.4em\relax Lawrence Erlbaum Associates,
  Inc, 1991, pp. 135--168.

\bibitem{citeulike:8845560}
P.~Ekman, \emph{{Basic Emotions}}.\hskip 1em plus 0.5em minus 0.4em\relax John
  Wiley \& Sons, Ltd, 2005, pp. 45--60.

\bibitem{Russell1991426}
J.~A. Russell, ``{Culture and the Categorization of Emotions},''
  \emph{Psychological Bulletin}, vol. 110, no.~3, pp. 426--450, 1991.

\bibitem{citeulike:9352253}
W.~Wundt, \emph{{Grundz\"{u}ge der physiologischen Psychologie}}.\hskip 1em
  plus 0.5em minus 0.4em\relax Leipzig: Engelmann, 1905.

\bibitem{citeulike:2239554}
M.~M. Bradley and P.~J. Lang, ``{Measuring emotion: the Self-Assessment Manikin
  and the Semantic Differential.}'' \emph{J. Behav. Ther. Exp. Psy.}, vol.~25,
  no.~1, pp. 49--59, 1994.

\bibitem{citeulike:9343328}
S.~Marsella, J.~Gratch, and P.~Petta, \emph{{Computational models of
  emotion}}.\hskip 1em plus 0.5em minus 0.4em\relax Oxford, UK: Oxford
  University Press, 2010, ch. 1.2, pp. 21--41.

\bibitem{citeulike:2642710}
J.~A. Russell and A.~Mehrabian, ``{Evidence for a three-factor theory of
  emotions},'' \emph{J. Res. Pers.}, vol.~11, no.~3, pp. 273--294, 1977.

\bibitem{Fontaine01122007}
J.~R.~J. Fontaine, K.~R. Scherer, E.~B. Roesch, and P.~C. Ellsworth, ``{The
  World of Emotions is not Two-Dimensional},'' \emph{Psychological Science},
  vol.~18, no.~12, pp. 1050--1057, 2007.

\bibitem{4959919}
Y.-H. Yang and H.~Chen, ``Music emotion ranking,'' in \emph{ICASSP 2009}, 2009,
  pp. 1657--1660.

\bibitem{EURECOM2361}
O.~Villon, ``{Modeling affective evaluation of multimedia contents: user models
  to Associate subjective experience, physiological expression and contents
  description},'' Ph.D. dissertation, Universit\'{e} de Nice - Sophia
  Antipolis, Nice, France, 2007.

\bibitem{6496208}
A.~Kazemzadeh, S.~Lee, and S.~Narayanan, ``Fuzzy logic models for the meaning
  of emotion words,'' \emph{IEEE Comput. Intell. Mag.}, vol.~8, no.~2, pp.
  34--49, 2013.

\bibitem{hunter2008mixed}
P.~G. Hunter, E.~G. Schellenberg, and U.~Schimmack, ``Mixed affective responses
  to music with conflicting cues,'' \emph{Cognition \& Emotion}, vol.~22,
  no.~2, pp. 327--352, 2008.

\bibitem{hunter2011misery}
P.~G. Hunter, E.~Glenn~Schellenberg, and A.~T. Griffith, ``Misery loves
  company: Mood-congruent emotional responding to music,'' \emph{Emotion},
  vol.~11, no.~5, pp. 1068--1072, 2011.

\bibitem{Desmet2003}
P.~Desmet, \emph{{Measuring emotion: development and application of an
  instrument to measure emotional responses to products}}.\hskip 1em plus 0.5em
  minus 0.4em\relax Norwell, MA, USA: Kluwer Academic Publishers, 2003, ch.~9,
  pp. 111--123.

\bibitem{Winoto20106086}
P.~Winoto and T.~Y. Tang, ``{The role of user mood in movie recommendations},''
  \emph{Expert Syst. Appl.}, vol.~37, no.~8, pp. 6086--6092, 2010.

\bibitem{citeulike:2772174}
J.~A. Russell, A.~Weiss, and G.~A. Mendelsohn, ``{Affect Grid: A single-item
  scale of pleasure and arousal},'' \emph{J. Pers. Soc. Psychol.}, vol.~57,
  no.~3, pp. 493--502, 1989.

\bibitem{Russell1980}
J.~A. Russell, ``{A circumplex model of affect.}'' \emph{J. Pers. Soc.
  Psychol.}, vol.~39, no.~6, pp. 1161--1178, 1980.

\bibitem{citeulike:8384089}
A.~Schaefer, F.~Nils, X.~Sanchez, and P.~Philippot, ``{Assessing the
  effectiveness of a large database of emotion-eliciting films: A new tool for
  emotion researchers},'' \emph{Cognition \& Emotion}, vol.~24, no.~7, pp.
  1153--1172, 2010.

\bibitem{citeulike:8758783}
D.~Watson and L.~A. Clark, ``{The PANAS-X: Manual for the Positive and Negative
  Affect Schedule--Expanded Form},'' 1994.

\bibitem{citeulike:3721917}
R.~Cowie, E.~Douglas-Cowie, S.~Savvidou, E.~Mcmahon, M.~Sawey, and
  M.~Schr\"{o}der. (2000) {feeltrace': an instrument for recording perceived
  emotion in real time}.

\bibitem{citeulike:7921301}
E.~Douglas-Cowie, R.~Cowie, and M.~Schr\"{o}der, ``{A New Emotion Database:
  Considerations, Sources and Scope},'' in \emph{ISCA Workshop (ITRW) on Speech
  and Emotion}, 2000, pp. 39--44.

\bibitem{5349526}
M.~Soleymani, J.~Davis, and T.~Pun, ``{A collaborative personalized affective
  video retrieval system},'' in \emph{Affective Computing and Intelligent
  Interaction}, 2009.

\bibitem{hanjalicxu:2005}
A.~Hanjalic and L.-Q. Xu, ``{Affective video content representation and
  modeling},'' \emph{IEEE Trans. Multimedia}, vol.~7, no.~1, pp. 143--154,
  2005.

\bibitem{1637510}
H.~L. Wang and L.-F. Cheong, ``{Affective understanding in film},'' \emph{IEEE
  Trans. Circuits Syst. Video Technol.}, vol.~16, no.~6, pp. 689--704, 2006.

\bibitem{4674668}
S.~Arifin and P.~Cheung, ``{Affective Level Video Segmentation by Utilizing the
  Pleasure-Arousal-Dominance Information},'' \emph{IEEE Trans. Multimedia},
  vol.~10, no.~7, pp. 1325--1341, 2008.

\bibitem{Xu:2008:HMA:1459359.1459457}
M.~Xu, J.~S. Jin, S.~Luo, and L.~Duan, ``Hierarchical movie affective content
  analysis based on arousal and valence features,'' in \emph{ACM int'l conf.
  Multimedia}, 2008, pp. 677--680.

\bibitem{citeulike:7927300}
M.~Soleymani, J.~J.~M. Kierkels, G.~Chanel, and T.~Pun, ``{A Bayesian framework
  for video affective representation},'' in \emph{Affective Computing and
  Intelligent Interaction}, 2009, pp. 1--7.

\bibitem{5571819}
G.~Irie, T.~Satou, A.~Kojima, T.~Yamasaki, and K.~Aizawa, ``{Affective
  Audio-Visual Words and Latent Topic Driving Model for Realizing Movie
  Affective Scene Classification},'' \emph{IEEE Trans. Multimedia}, vol.~12,
  no.~6, pp. 523--535, 2010.

\bibitem{citeulike:8161572}
H.~Joho, J.~Staiano, N.~Sebe, and J.~Jose, ``{Looking at the viewer: analysing
  facial activity to detect personal highlights of multimedia contents},''
  \emph{Multimed. Tools App's}, vol.~51, no.~2, pp. 505--523, 2010.

\bibitem{citeulike:8638639}
R.~M. Teixeira, T.~Yamasaki, and K.~Aizawa, ``{Determination of emotional
  content of video clips by low-level audiovisual features},'' \emph{Multimed.
  Tools App's}, pp. 1--29, 2011.

\bibitem{DBLP:conf/eccv/DemartyPGS12}
C.-H. Demarty, C.~Penet, G.~Gravier, and M.~Soleymani, ``A benchmarking
  campaign for the multimodal detection of violent scenes in movies,'' in
  \emph{ECCV Workshops (3)}, ser. LNCS, A.~Fusiello \emph{et~al.}, Eds., vol.
  7585.\hskip 1em plus 0.5em minus 0.4em\relax Springer, 2012, pp. 416--425.

\bibitem{Joho:2009:EFE:1646396}
H.~Joho, J.~M. Jose, R.~Valenti, and N.~Sebe, ``{Exploiting facial expressions
  for affective video summarisation},'' in \emph{ACM Int'l Conf. Image and
  Video Retrieval}, 2009.

\bibitem{citeulike:6922056}
J.~J.~M. Kierkels, M.~Soleymani, and T.~Pun, ``Queries and tags in affect-based
  multimedia retrieval,'' in \emph{IEEE int'l conf. Multimedia and Expo}, 2009,
  pp. 1436--1439.

\bibitem{Soleymani:2009:IJSC}
M.~Soleymani, G.~Chanel, J.~J.~M. Kierkels, and T.~Pun, ``{Affective
  Characterization of Movie Scenes Based on Content Analysis and Physiological
  Changes},'' \emph{Int'l J. Semantic Computing}, vol.~3, no.~2, pp. 235--254,
  2009.

\bibitem{mahnob_HCI}
M.~Soleymani, J.~Lichtenauer, T.~Pun, and M.~Pantic, ``A multimodal database
  for affect recognition and implicit tagging,'' \emph{IEEE Trans. Affective
  Computing}, vol.~3, pp. 42--55, 2012.

\bibitem{citeulike:8017077}
J.~Rottenberg, R.~D. Ray, and J.~J. Gross, \emph{{Emotion elicitation using
  films}}, ser. Series in affective science.\hskip 1em plus 0.5em minus
  0.4em\relax Oxford University Press, 2007, pp. 9--28.

\bibitem{citeulike:775844}
P.~Ekman \emph{et~al.}, ``{Universals and cultural differences in the judgments
  of facial expressions of emotion.}'' \emph{J. Pers. Soc. Psychol.}, vol.~53,
  no.~4, pp. 712--717, 1987.

\bibitem{citeulike:8791184}
R.~Plutchik, \emph{{A general psychoevolutionary theory of emotion}}.\hskip 1em
  plus 0.5em minus 0.4em\relax New York: Academic press, 1980, pp. 3--33.

\bibitem{Soleymani:CSE2010}
M.~Soleymani and M.~Larson, ``Crowdsourcing for affective annotation of video:
  Development of a viewer-reported boredom corpus,'' in \emph{Workshop on
  Crowdsourcing for Search Evaluation, SIGIR 2010}, Geneva, Switzerland, 2010.

\bibitem{web_studies}
U.-D. Reips, ``\BIBforeignlanguage{English}{The web experimental psychology
  lab: Five years of data collection on the internet},''
  \emph{\BIBforeignlanguage{English}{Beh. Res. Meth. Instr. Comp.}}, vol.~33,
  no.~2, pp. 201--211, 2001.

\bibitem{citeulike:2686454}
A.~Kittur, E.~H. Chi, and B.~Suh, ``{Crowdsourcing user studies with Mechanical
  Turk},'' in \emph{ACM SIGCHI conf. Human factors in computing systems (CHI)},
  2008, pp. 453--456.

\bibitem{citeulike:8847316}
J.~D. Laird, \emph{{Feelings: The Perception of Self}}, 1st~ed.\hskip 1em plus
  0.5em minus 0.4em\relax USA: Oxford University Press, 2007.

\bibitem{paolacci2010}
G.~Paolacci, J.~Chandler, and P.~G. Ipeirotis, ``{Running Experiments on Amazon
  Mechanical Turk},'' \emph{Judgm Decis Mak.}, vol.~5, no.~5, pp. 411--419,
  2010.

\bibitem{Quirin2009500}
M.~Quirin, M.~Kaz\'{e}n, and J.~Kuhl, ``{When Nonsense Sounds Happy or
  Helpless: The Implicit Positive and Negative Affect Test (IPANAT)},''
  \emph{J. Pers. Soc. Psychol.}, vol.~97, no.~3, pp. 500--516, 2009.

\end{thebibliography}
\vspace{-40pt}
 \begin{IEEEbiography}[{\includegraphics[width=1in]{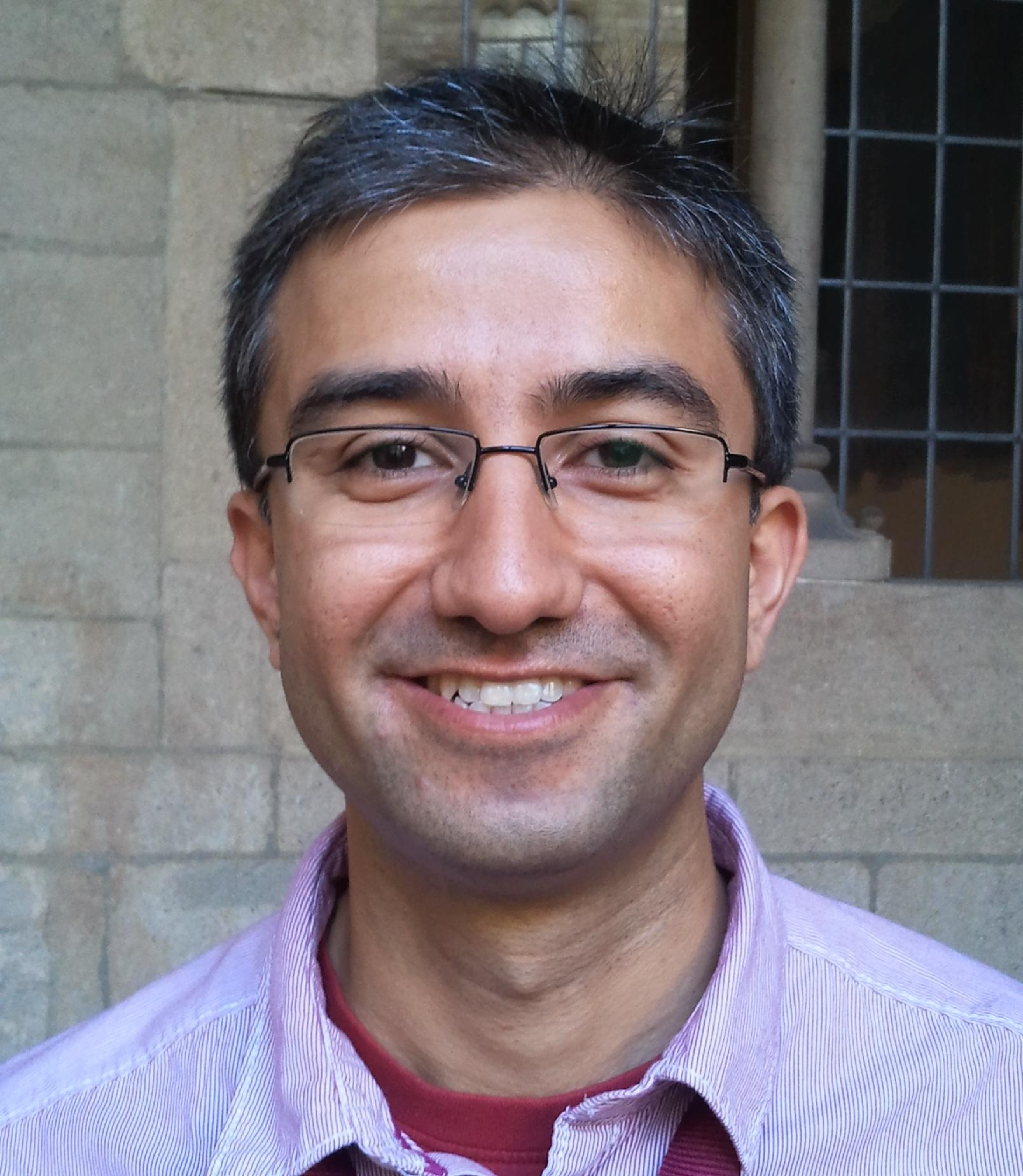}}]{Mohammad Soleymani}
 is a Marie Curie Fellow at Imperial College London, where he conducts research on sensor-based and implicit emotional tagging.  Soleymani received his PhD in computer science from the University of Geneva, Switzerland in 2011. He has worked extensively on assessing emotional reactions in response to video content and developing multimedia techniques to predict these reactions.  He  has served as a special session chair, program committee member and reviewer for multiple conferences and workshops including ACM ICMR, ACM MM, ACM ICMI, IEEE SMC, and IEEE ICME.
 \end{IEEEbiography}
 \vspace{-20pt}
 \begin{IEEEbiography}[{\includegraphics[width=1in]{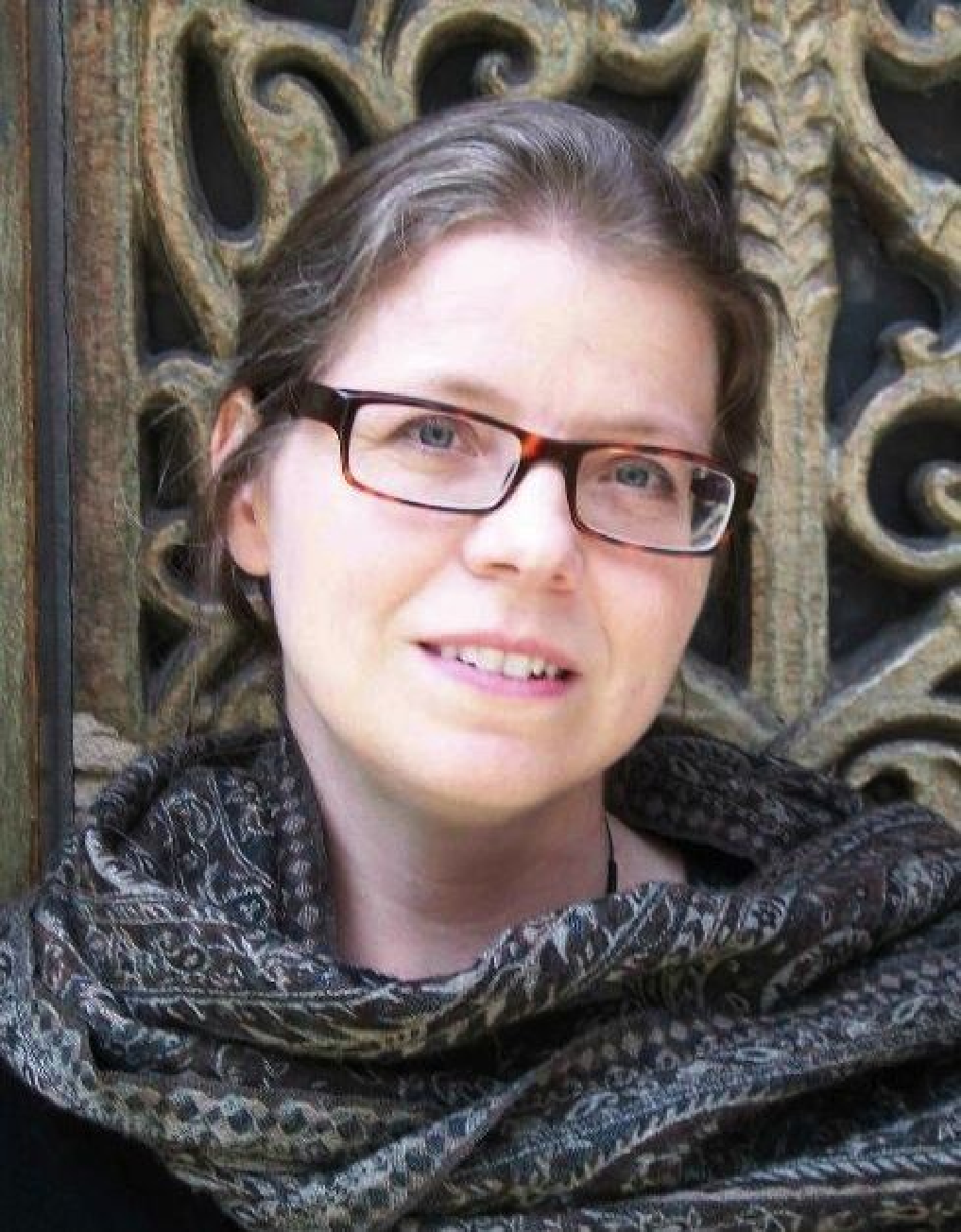}}]{Martha Larson}
 holds a MA and PhD in theoretical linguistics from Cornell University and a BS in Mathematics from the University of Wisconsin. Her research interest and expertise lie in the area of speech- and language-based techniques for multimedia information retrieval. She is co-founder of the MediaEval Multimedia Benchmark and has served as organizer of a number of workshops in the areas of spoken content retrieval and crowdsourcing. She has authored or co-authored over 100 publications. Currently, Dr. Larson is assistant professor in the Multimedia Information Retrieval Lab at Delft University of Technology. Before coming to Delft, she researched and lectured in the area of audio-visual retrieval at Fraunhofer IAIS and at the University of Amsterdam.
 \end{IEEEbiography}
  \vspace{-20pt}

 \begin{IEEEbiography}[{\includegraphics[width=1in]{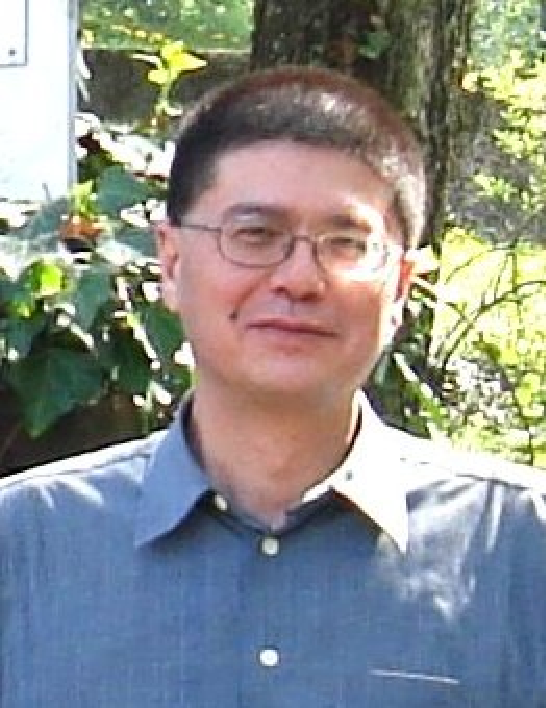}}]{Thierry Pun}

Thierry Pun is head of the Computer Vision and Multimedia Laboratory, full professor at the Computer Science Department, University of Geneva, Switzerland (http://cvml.unige.ch). His current research interests, related to affective computing and multimodal interaction, concern physiological and behavioural signals analysis for affective state assessment, affective gaming and learning, affect in social media, brain-computer interaction, multimodal interfaces for blind users and for the elderly. He has authored or co-authored over 300 full papers as well as eight patents. He is in the editorial boards of the International Journal on Image \& Video Processing (Springer), and Advances in Multimedia (Hindawi). He was one of the general chairs of ACII - Affective Computing and Intelligent Interaction 2013 in Geneva.
 \end{IEEEbiography}
  \vspace{-20pt}

 \begin{IEEEbiography}[{\includegraphics[width=1in]{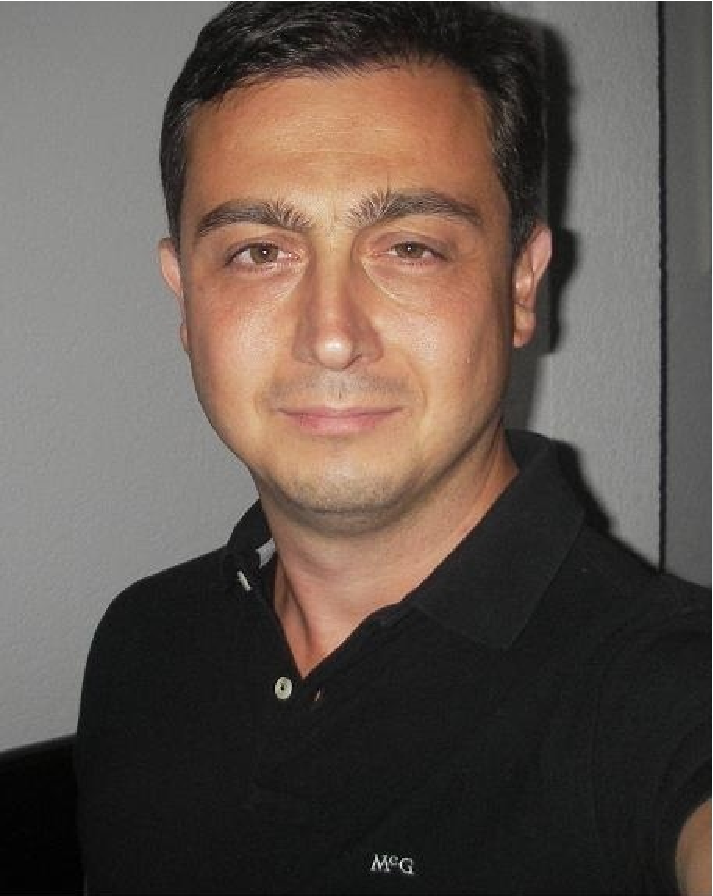}}]{Alan Hanjalic} 
Alan Hanjalic is a professor and head of the Multimedia Signal Processing Group at the Delft University of Technology, The Netherlands. His research focus is on multimedia information retrieval, recommender systems and social media analytics. Prof. Hanjalic is an elected member of the IEEE Technical Committee on Multimedia Signal Processing and an appointed member of the Steering Committee of the IEEE Transactions on Multimedia. He has been a member of Editorial Boards of five scientific journals in the multimedia field, including the IEEE Transactions on Multimedia, the IEEE Transactions on Affective Computing and the International Journal of Multimedia Information Retrieval. He has also held key positions in the organizing committees of leading multimedia conferences, including the ACM Multimedia, ACM ICMR and IEEE ICME.
\end{IEEEbiography}
%\end{IEEEbiography}


%\newpage


%\vfill


\end{document}